\documentclass[journal,hideappendix]{vgtc}        

\onlineid{1729}

\vgtccategory{Research}

\vgtcpapertype{Applications}

\title{ \sysName{}: Guiding the Creation of Multi-agent Workflows with Design Space Visualization as a Thinking Scaffold}

\author{%
    \authororcid{Pan Hao}{0009-0006-4473-5764},
  \authororcid{Dongyeop Kang}{0000-0002-9021-1789},
  Nicholas Hinds, and 
  \authororcid{Qianwen Wang}{0000-0002-1825-0097}
}

\authorfooter{
  	\item Pan Hao, Dongyeop Kang, Nicholas Hinds, and Qianwen Wang are with the University of Minnesota, Twin Cities, MN, USA. E-mail: \{pan00342, dongyeop, hinds084, qianwen\}@umn.edu
}

\abstract{%
  Multi-agent workflows have become an effective strategy for tackling complicated tasks by decomposing them into multiple sub-tasks and assigning them to specialized agents.
  However, designing optimal workflows remains challenging due to the vast and intricate design space. 
  Current practices rely heavily on the intuition and expertise of practitioners, often resulting in design fixation or an unstructured, time-consuming exploration of trial-and-error.
  To address these challenges, this work introduces \sysName{}, an interactive visualization tool to facilitate the creation of multi-agent workflow through i) a structured visual exploration of the design space and ii) in-situ guidance informed by established design patterns.
  Based on formative studies and literature review, \sysName{} organizes the workflow design process into three hierarchical levels (\ie, task planning, agent assignment, and agent optimization), ranging from abstract to concrete. 
  This structured visual exploration enables users to seamlessly move from high-level planning to detailed design decisions and implementations, while comparing alternative solutions across multiple performance metrics. Additionally, drawing from established workflow design patterns, \sysName{} provides context-aware, in-situ suggestions at each level as users navigate the design space, enhancing the workflow creation process with practical guidance.
  Use cases and user studies demonstrate the usability and effectiveness of \sysName{}, while also yielding valuable insights into how practitioners explore design spaces and leverage guidance during workflow development.
}


\teaser{
  \centering
  \includegraphics[width=\linewidth, alt={A view of a city with buildings peeking out of the clouds.}]{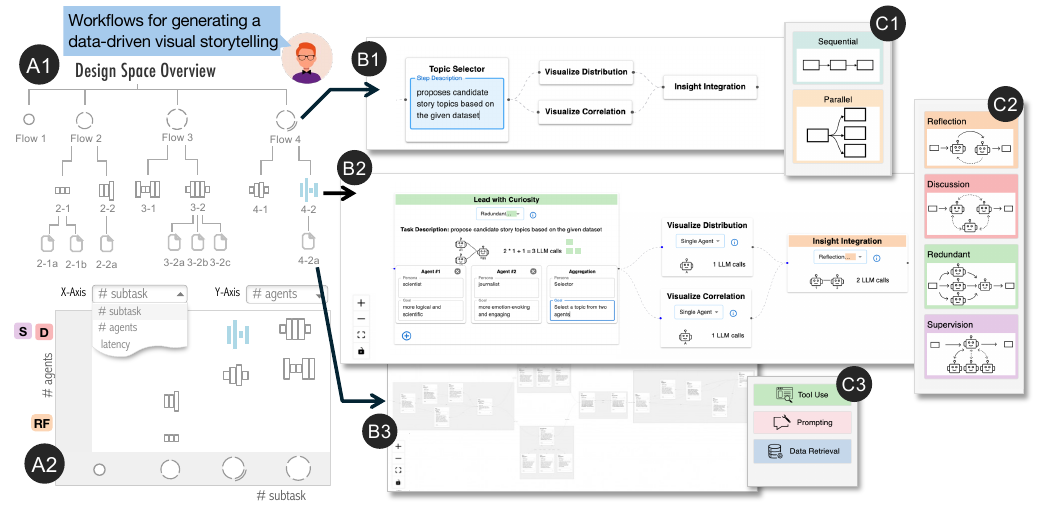}
  \vspace{-2em}
  \caption{%
  \sysName{} facilitates the creation of multi-agent workflows through structured, guided visual exploration of the design space. 
  This is achieved by coordinating a hierarchical tree view (A1) and a scatter plot (A2), both of which use a novel glyph design to represent each workflow’s computational cost and level of abstraction. 
  Users can select a workflow for detailed inspection in the Canvas View (B1-B3), which supports semantic zooming that reveals more abstract representations greater detail within more abstract representations depending on the zoom level. 
  Additionally, in-situ suggestions based on well-established design patterns are provided to guide users as they navigate the design space (C1-C3).
  }
  \label{fig:teaser}
}

\graphicspath{{figs/}{figures/}{pictures/}{images/}{./}} 

\usepackage{tabu}                      
\usepackage{booktabs}                  
\usepackage{lipsum}                    
\usepackage{mwe}                       

\usepackage{mathptmx}                  

\newcommand{\ie}{\textit{i.e.}}
\newcommand{\eg}{\textit{e.g.}}
\newcommand{\etal}{\textit{et al.}}
\newcommand{\sysName}{\textsc{FlowForge}}

\newcommand{\qianwen}[1]{#1}

\usepackage{duckuments}
\usepackage{enumitem}
\usepackage{multirow}

\usepackage{flushend}
\begin{document}


\firstsection{Introduction}

\maketitle

Recently, multi-agent workflows have emerged as a promising solution for complex tasks by decomposing them into subtasks, each handled by specialized LLM agents that collaboratively reason and execute tasks \cite{anthropic2024agents, aws2024orchestrator}. 
These multi-agent workflows have demonstrated remarkable capabilities across various applications, such as software development, societal simulation, and even scientific discovery~\cite{guo2024survey, park2023generative}. 
Notably, Google's AI co-scientist \cite{gottweis2025towards}, a multi-agent LLM workflow system for scientific discovery, successfully uncovered a novel gene transfer mechanism, reproducing unpublished experimental findings \cite{penades2025ai}.

However, designing effective multi-agent workflows remains a considerable challenge due to the vast design space.
\qianwen{AI researchers and practitioners must balance critical trade-offs between performance, latency, and computational cost\cite{Grunde-McLaughlin2024crowdsourcing, liu2024cataloque}.
}
Despite recent advancements in automatic task decomposition and agent collaboration (\eg, Manus AI~\cite{ManusAI}), the over-reliance on LLMs for workflow generation not only compromises transparency and controllability but also often results in suboptimal solutions~\cite{cemri2025multi}.
Rather than relying on fully automated methods, \qianwen{AI researchers and practitioners} typically draw on their own intuition and domain expertise, leveraging well-established human problem-solving strategies~\cite{liu2024cataloque, Grunde-McLaughlin2024crowdsourcing}. 

Current studies predominantly focus on implementation support, \qianwen{helping users (\ie, AI researchers and practitioners) translate existing workflow ideas into executable systems}, but offer limited assistance for the exploratory design processes \cite{wu2023autogen, langgraph}. 
However, people's problem-solving approaches are heavily influenced by the available tools, which makes certain actions more accessible while rendering others difficult or impractical \cite{heidegger1977question}.
In the context of multi-agent workflows, the affordances of existing tools have similarly shaped common practices in ways that discourage broad exploration.
Users often begin with a rough idea of the workflow and move directly to implementation, a process that tends to bypass divergent thinking and leads to several limitations:
First, early commitment to a specific workflow structure can restrict exploration of alternative solutions that might perform better within the broader design space.
This contradicts the widely accepted divergence-before-convergence principle~\cite{buxton2010sketching}, leading to fixation on suboptimal designs.
Second, by starting with complete implementations, users skip over abstraction layers in the design space. When performance issues arise, they tend to make superficial adjustments (\eg, tweaking agent prompts) rather than revisiting fundamental design decisions (\eg, task decomposition strategies) where greater improvements could be made.

Recent studies attempted to address this gap by introducing design patterns specifically tailored to multi-agent workflows.
Design patterns represent reusable, adaptable solutions proven effective across various contexts.
For example, Liu \etal. \cite{liu2024cataloque} identified 18 agent design patterns (\eg, self-reflection, voting) through an extensive review of 57 multi-agent studies.
Grunde-McLaughlin \etal. \cite{Grunde-McLaughlin2024crowdsourcing} summarized five architecture workflow building blocks (\ie, sequential, branching, redundant, dynamic, communicative) by borrowing lessons about task decomposition and validation from crowdsourcing.
Even though these studies provide valuable guidance, these patterns remain disconnected from practical tools and environments, requiring \qianwen{AI researchers and practitioners} to actively seek out, efficiently learn, and manually apply them during the design process.

To address these challenges, \qianwen{we propose \sysName{}, an interactive visualization tool to facilitate AI researchers and practitioners in creating effective multi-agent workflows (\cref{fig:teaser}).} 
\sysName{} not only provides an intuitive visual interface for building workflows, but, more importantly, actively scaffolds users' design thinking by explicitly visualizing the design space and providing in-situ design guidance. 
Based on formative studies and literature review, \sysName{} organizes the workflow design process into three levels (\ie, task planning, agent assignment, and agent optimization), progressing from abstract concepts to concrete implementation. 
A structured visual exploration of the design space is enabled through the coordination of a hierarchical tree view and a scatter plot. Within this coordinated visualization, we introduce a novel glyph design that encodes both the abstraction level and computational cost of each solution. 
Users can select any solution for closer inspection in the Canvas View, where it is visualized as a node-link diagram with semantic zooming to reveal increasing detail as users zoom in.
Additionally, drawing from established workflow design patterns, \sysName{} provides contextually relevant in-situ suggestions at each level as users navigate the design space, illustrating current practice to the workflow creation process.
Evaluations showed that \sysName{} enabled users to create workflows more efficiently, explore a wider range of design alternatives, and produce higher-quality outputs compared to the baseline system.

Main contributions of this paper include:
\begin{itemize}[noitemsep, leftmargin=*, nosep]
    \item A structured characterization and visualization of the design space for multi-agent workflows, augmented with in-situ suggestions of relevant design patterns.
    \item \sysName{}, an interactive visualization tool that guides users in designing and creating multi-agent workflows.
    \item Case studies and user studies demonstrating the effectiveness and usability of the proposed approach, along with insights into how users leverage guidance to explore and refine workflow designs. 
\end{itemize}

\section{Related Work}

\subsection{Creating Multi-agent Workflows}
Existing research on multi-agent workflow creation can be categorized into three main areas: workflow authoring frameworks, empirical design guidelines, and automated generation methods.

Workflow \textbf{authoring frameworks} provide programming libraries or visualization tools for simplifying multi-agent workflow creation. 
Programming frameworks, such as LangGraph \cite{langgraph} and AutoGen \cite{wu2023autogen}, offer APIs that allow users to efficiently define agent roles and manage inter-agent communication. 
To make workflow creation more accessible to non-programmers, no-code platforms, such as Rivet \cite{rivet}, Vellum \cite{vellum}, and AutoGen Studio \cite{dibia2024autogenstuido}, offer visual interfaces for designing and testing complex workflows. 
Users can define agents as compact cards and construct workflows using a drag-and-drop interface.
Additionally, interactive tools, like AI Chains and Prompt Chainer \cite{wu2022prompt}, streamline the process of chaining prompts via interactive visualizations.
However, these studies primarily assist in implementing workflows that users have already conceptualized and still heavily rely on user expertise for workflow design (\eg, task planning and orchestrating communication among agents).

To mitigate the complexity in designing workflows, recent studies have derived \textbf{design patterns and guidelines} from systematic investigation of existing workflow implementations~\cite{liu2024cataloque, Grunde-McLaughlin2024crowdsourcing, guo2024survey, zhou2025multi}. 
For example, Liu \etal \cite{liu2024cataloque} identified 18 design patterns (\eg, self-reflection, voting, Role-based cooperation) across various scenarios.
Grunde-McLaughlin \etal \cite{Grunde-McLaughlin2024crowdsourcing} summarized and tested five workflow building blocks (\ie, sequential, branching, redundant, dynamic, communicative) by drawing inspiration from crowdsourcing methodologies and study designs. 
Meanwhile, industry guidelines from major technology companies have increasingly emerged, outlining common practices used in the community, such as group chat, debate, and reflection \cite{microsoft2024patterns, microsoft2024conversation, anthropic2024agents}.

\textbf{Automated methods} offer another research direction, focusing on reducing user workload by automatically generating optimized workflow designs.
One prevalent approach involves introducing a ``supervisor'' agent that decomposes tasks and orchestrates agent collaboration. 
For instance, ADAS \cite{hu2024ADAS} employs a meta-agent to generate and refine reusable code snippets, assembling them into complete workflows.
Another approach models the search for suitable workflows as an optimization problem that can be solved computationally~\cite{zhuge2024gptswarm, zhang2024aflow, zhou2025multi}.
For example, GPTSwarm \cite{zhuge2024gptswarm} formulates multi-agent workflow construction as an optimization over computational graphs, with agents as nodes and communication channels as edges.
While these methods simplify the task orchestration, they often sacrifice transparency, reliability, and controllability, which are crucial in critical or high-stakes scenarios.

As a result, existing studies tend to fall into two extremes, either fully relying on a user's own knowledge or entirely automating the workflow creation. 
Our proposed method aims to strike a balance by embedding design space exploration and empirical design guidance directly into the workflow-building environment.






\subsection{Visualizing Design Space}
A design space encompasses the conceptual realm of possible solutions for a given problem and serves as a fundamental framework for structured exploration in various design processes, including machine learning architecture design~\cite{wang2019atmseer, yuan2022visual}, creative writing~\cite{suh2024luminate}, and image generation~\cite{brade2023promptify, feng2023promptmagician}. 
Effective visualization of design spaces helps users navigate and interpret the numerous solutions by clearly representing their dimensions and abstraction levels.

\textbf{Dimensions} represent key parameters that define and distinguish various possible solutions within a design space, and can be used to effectively categorize different design choices.
For example, Luminate~\cite{suh2024luminate} visualizes the design space of creative writing via dimensions such as genres, personality, story tone.
Similarly, PromptMagician~\cite{feng2023promptmagician} and Promptify~\cite{brade2023promptify} structure the design space of text-to-image generation by incorporating object- and style-related dimensions. 
Various visualization techniques, including scatter plots~\cite{suh2024luminate, brade2023promptify}, tree diagrams~\cite{suh2023sensecape, wang2019visual}, and parallel coordinates~\cite{yuan2022visual}, have been employed to represent these dimensions and organize various solutions.

\textbf{Levels} of abstraction further facilitate design space navigation by hierarchically structuring solutions from high-level concepts to fine-grained specifics.
Different visualization techniques are often employed to effectively represent each abstraction layer.
For example, ATMSeer \cite{wang2019atmseer} visualizes the design space of machine learning models through three hierarchical levels (\ie, algorithm type, hyper-partition, and hyper-parameter) using histogram, bar chart, and scatter plot, respectively.
Meanwhile, studies have used focus + context technique to allow users to view both an abstract overview and detailed information simultaneously \cite{yuan2022visual, wang2019visual}. 
DNN Genealogy\cite{wang2019visual} dynamically displaying solutions as either text labels or glyphs based on the calculated degree of interest.
Semantic zooming is another technique for navigating hierarchical design spaces, dynamically adapting visual representations based on zoom level. 
For example, when visualizing the design space of LLM prompts, abstract level  (zoomed out) represents solutions as simple points, whereas detailed level(zoomed in) reveals full text outputs or generated images~\cite{suh2024luminate, feng2023promptmagician, brade2023promptify}.

Our work builds upon these prior visualization methods, uniquely adapting them to scaffold user cognition and provide in-situ support during multi-agent workflow design. 

\section{Designing \sysName{}}



\subsection{Understanding Design Challenges}
\label{subsec:understanding}

To identify the current challenges in designing multi-agent workflows, we conducted a formative study involving four AI researchers (E1-E4) with extensive experience in multi-agent systems.
\qianwen{These participants represent our target user base of AI researchers and practitioners who design, development, and evaluate multi-agent workflows for solving complex tasks.}
Participants were recruited through personal networks and snowball sampling. All participants had led at least one major project about multi-agent LLM systems.
One expert is the author of this paper while others are not.
Each session consisted of a 45-60 minute semi-structured interview conducted over Zoom.
Participants were asked to talk about their current practice in multi-agent workflows. 
They also reflected on the difficulties in constructing multi-agent workflows, and the limitations of the current tools.
With consent from the participants, the interviews were recorded, transcribed, and thematically analyzed.
We identified four key challenges:

\begin{enumerate}[leftmargin=*, label=\textbf{C.\arabic* }, labelsep=0pt]
    \item \textbf{Vast Design Space.}
    \label{challenge:vast}
    All participants emphasized designing a multi-agent workflow involves navigating a vast and complex design space.
    Designers must make decisions across several dimensions: how to decompose a task, whether a task should be handled by a single agent or multiple agents, how agents should communicate, and how to configure individual agent behavior.
    Despite their experience, participants relied heavily on intuition, past examples, and human problem-solving analogies.
    For complex tasks, they noted that relying on ``supervisor'' agents for orchestration was often insufficient, and human oversight remained critical. 
    The vast size and high complexity of design space made exploration difficult without structured support. 
    \item \textbf{Unstructured Trial-and-error Exploration.} 
    \label{challenge:unstructured}
    Workflow design often follows an ad hoc trial-and-error process.
    Participants described iterating on ideas after failures, often tweaking small elements (\eg, prompts), mid-level elements (\eg, conversation rounds), or larger architectural components (\eg, switching from hierarchical to sequential designs). 
    However, without a structured framework, this process was time-consuming, frustrating, and frequently unguided. 

   \item \textbf{Design Fixation.}
   \label{challenge:fixation}
   Practitioners reported that they often begin by implementing a single workflow first and then iterating to refine it. 
   Due to the complexity of implementing and fine-tuning a workflow, practitioners are often hesitant to explore alternatives. Most of their exploration were reactive in response to failures, rather than proactively seeking better designs.
   Once a runnable version of the workflow is achieved, further modifications were usually incremental, leading to premature convergence and missed opportunities for better-performing designs.
   

    \item \textbf{Navigating Trade-offs between Performance Metrics.}
    \label{challenge:metrics}
    Navigating multi-agent workflows often requires balancing conflicting priorities, such as computational cost, response latency, creativity, and accuracy.
    These trade-offs are subjective and context-dependent, such as generating fast forward videos for research papers (E1) and writing politics essays (E2), involving multiple design considerations.    
    For example, E3 commented that \textit{``latency was the biggest concern''} in one previous project, driving all design decisions. 
    At the same time, it is challenging to account for different metrics simultaneously during the design phase.
    Some considerations (\eg, practical costs) often become apparent only after implementation, at which point practitioners may be reluctant to make major changes.

\end{enumerate}


\begin{figure}
    \centering
    \includegraphics[width=0.9\linewidth]{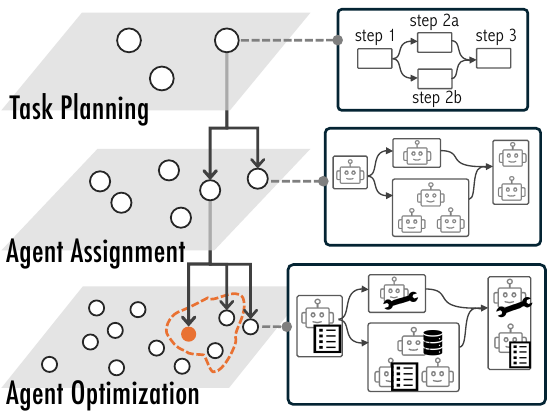}
    \caption{The design space of multi-agent workflows can be conceptualized across three hierarchical levels. Current frameworks primarily operate at the most granular level, which often leads to design fixation (the orange dot) rather than encourage exploration of the full solution space.}
    \label{fig:levels}
\end{figure}

\begin{table*}[tb]
  \caption{Design Patterns for Multi-Agent Workflows supported in \sysName{}. 
}
  \label{tab:design_patterns}
  \small
	\centering%
    
  \begin{tabu}{%
        p{0.05cm}|
	    p{2.4cm}%
        p{5.2cm}%
        p{6.0cm}
        p{1.9cm}
	}
  \toprule
     
    \multicolumn{2}{c}{ Pattern }
    & Definition 
    & Example  
    &  References    
    \\
  \midrule
  
  \multirow{2}*{\centering\rotatebox[origin=c]{90}{\textbf{Level-1}}}
   &Sequential \newline
  \includegraphics[width=2cm]{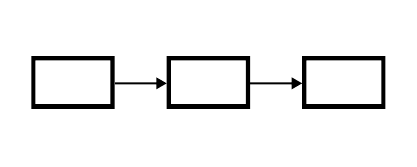}
  & The task can be decomposed into a sequence of steps, which are executed sequentially and completed in an orderly progression. 
  &  Debugging visualizations via first interpreting code, identifying issues, and then writing code \cite{wu2022chains}. 
  &\cite{wu2022chains} \cite{anthropic2024agents} \cite{crewai} \cite{wu2023crowdsoucingllms} \cite{qian2024chatdev}
  \\

    \cline{2-5}
    &Parallel \newline
  \includegraphics[width=2cm]{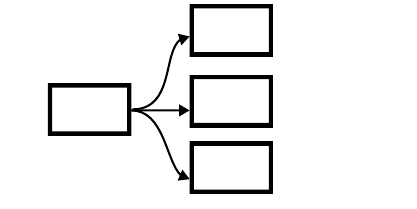}
  & The task can be divided into multiple subtasks that can be parallelized for potential speedup.
  & When generating a travel plan, ideating various vacation activities simultaneously \cite{wu2022prompt}. 
  &\cite{wu2022chains} \cite{anthropic2024agents} \cite{prasad2023adapt} \cite{wu2023crowdsoucingllms} \cite{qian2024chatdev}
    \\
  \hline

  \multirow{4}*{\centering\rotatebox[origin=c]{90}{\textbf{Level-2}}}
  &Reflection \newline
  \includegraphics[width=2.2cm]{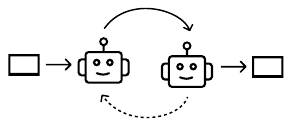}
  &  One agent generates and refines a response while another evaluates it, repeating the process until a predefined criterion (quality or number of iterations) is met.
  Often used to improve accuracy.
  & One agent generates code, which a reviewer then assesses for quality and returns feedback for adjustment \cite{wu2023autogen}.
  &\cite{anthropic2024agents}\cite{shinn2023reflexion} \cite{microsoft2024patterns}\cite{liu2024cataloque}\cite{shinn2023reflexion}\cite{zhuge2024gptswarm}\cite{hu2024ADAS}\cite{AG2_2024}\cite{zhang2024optimize}
  \\
  
    \cline{2-5}
  &Redundant \newline
  \includegraphics[width=2.2cm]{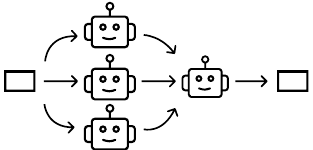}
  & 
  Multiple agents simultaneously respond to the same input while adopting different personas or perspectives, generating diverse outputs. 
  Often used to improve robustness or creativity.
  & 
  Four agents in roles of physics, chemistry, biology, and general science experts attempt to solve graduate-level questions in science individually, then their answers are aggregated~\cite{hu2024ADAS}. 
  &\cite{langgraph2024workflows}\cite{anthropic2024agents} \cite{Grunde-McLaughlin2024crowdsourcing}\cite{niu2025flowmodular}\cite{hu2024ADAS}
  \\

  
  \cline{2-5}
  &Supervision \newline
  \includegraphics[width=2.2cm]{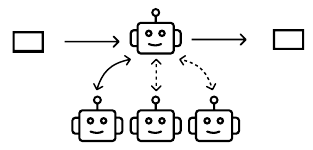}
  & A hierarchical structure where a central agent dynamically decides which worker agents to act based on the input. Often used for complicated tasks that require diverse expertise and careful quality control. 
  & 
  A supervisor agent directs user inquiries to specialized agents based on intent—such as technical support, refund and return processing, or handling complaints~\cite{aws2024orchestrator}. 
  &\cite{anthropic2024agents}\cite{langgraph2024multiagent} \cite{microsoft2024patterns}\cite{aws2024orchestrator}\cite{AG2_2024}\cite{liu2024cataloque}\cite{guo2024survey}\cite{crewai}\cite{aws2024orchestrator}\cite{zhang2025chainbuddy}\cite{wu2022prompt}
  \\
  \cline{2-5}
  &Discussion \newline
  \includegraphics[width=2.2cm]{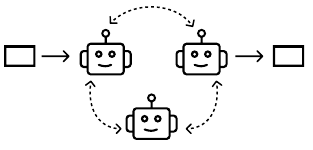}
  & A group of agents engage in a conversation, where they take turns speaking in a round-robin, random manner, or simultaneously. Often used to increase creativity for open-ended tasks.
  & Three agents in roles of news author, critic, and psychologist take turns to generate their response based on the conversation to evaluate open-ended question answers~\cite{chan2023chateval}. 
  &\cite{AG2_2024}\cite{chan2023chateval} \cite{liang2024debatedivergent} \cite{du2023mathdebate}\cite{microsoft2024patterns}\cite{guo2024survey}\cite{liu2024cataloque}
  \\
  \hline

  \multirow{1}*{\centering\rotatebox[origin=c]{90}{\textbf{Level-3}}}
  &Single Agent \newline
  \includegraphics[width=2.0cm]{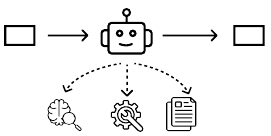}
  & The basic building block of multi-agent workflows, a single LLM enhanced with augmentations such as data retrieval, tools, prompting, and memory. 
  &  A single agent augmented with Web search tool queries for question-answering \cite{nakano2022webgptbrowser}. A basic module with diverse augmentations to enhance individual ability and versatility. 
  &\cite{langgraph2024multiagent} \cite{anthropic2024agents} \cite{hu2024ADAS} \cite{wu2023autogen} \cite{liu2024cataloque} \cite{guo2024survey} \cite{crewai}\cite{li2023camel}\cite{AG2_2024}\cite{nakano2022webgptbrowser}
    \\
  \hline
\end{tabu}%
\end{table*}

\subsection{Design Space of Multi-agent Workflows}
To address these challenges, \sysName{} summarizes and visualizes the design space of multi-agent workflows, offering users a structured overview of various possible solutions. 
This subsection introduces the three key components of this design space: \textbf{design dimensions}, \textbf{reusable design patterns}, \textbf{hierarchical abstraction levels}, grounded in both user study and recent literature.
While these components are not exhaustive and are likely to evolve with ongoing research, they provide a strong foundation for demonstrating the usefulness of visualizing design space and integrating established design patterns.

\subsubsection{Design Dimensions}
\label{subsubsec:dimensions}
Dimensions are key factors that define and distinguish various workflows within the design space, such as the running time, the number of agents, and the rate of hallucination.
Certain key dimensions are common across various tasks, while others are specific to particular applications.
For example, dimensions such as accuracy and computational efficiency are universally relevant to most tasks, as practitioners generally seek to optimize them regardless of application.
In contrast, specialized dimensions like the degree of dynamic updating are important for specific contexts. Our formative study revealed that while experts (E1-E4) unanimously rejected fully automated workflow generation, they sometimes incorporated components with dynamic updating capabilities (\eg, agents that select subsequent agents based on output analysis) in particular scenarios where adaptability was essential.
Dimensions can also be highly granular, addressing specific characteristics like output length, formatting elements (\eg, bullet points), or the presence of certain keywords. They can influence design choices across the design process, such as reducing the number of agents to lower computational costs.
Since these design dimensions can be very task-specific, there is not an exclusive list of all possible dimensions.
\subsubsection{Design Patterns}
\label{subsec:}
Design patterns are reusable, generalizable strategies for structuring workflows that emerge from repeated adaptation across numerous existing implementations. 
A design patterns is often associated with optimizing specific performance dimensions, \eg, reflection based design pattern can improve accuracy.


\qianwen{Even though existing surveys provided valuable insights about workflow design, they either focus on one aspect of the design process (\eg, task planning \cite{Grunde-McLaughlin2024crowdsourcing}) or overlook the high-level abstraction within the design space and focus on implementation details~\cite{guo2024survey}.
Therefore, we extended existing workflow surveys \cite{liu2024cataloque, Grunde-McLaughlin2024crowdsourcing, guo2024survey} and analyzed 43 multi-agent systems drawn from papers, blog posts, framework documentation, via keyword-based searches (\eg, ``multi-agent'', ``LLM workflows'', ``agent orchestration'').}
Using axial coding, we grouped recurring structures into design patterns.
While terminology varies across sources (\eg, ``Group Chat'' vs ``Discussion''), we adopted the most commonly used terms in our dataset for consistency. 
A full summary of these patterns is provided in \autoref{tab:design_patterns}.
Although our collection does not capture every possible structure (\eg, dynamic prompt-based feedback \cite{hu2024ADAS}), it provides a robust foundation that effectively demonstrates the value of visualizing the design space and integrating established patterns into workflow creation processes.
The detailed analysis process is included in the project web page.

\subsubsection{Hierarchical Abstraction Levels}
\label{subsubsec:level}
\qianwen{We model the design space as three levels of abstraction (\cref{fig:levels}).
This approach follows the established practices in design-related fields that employ layered abstractions to manage complexity and structure the design process \cite{munzner2009nested, wang2022extending}. 
To construct each abstraction layer, we employ axial coding to organize the design patterns documented in our literature review, alongside the important design decisions and challenges identified through our formative study. This coding process groups related concepts into three hierarchical levels that progressively address the workflow development complexity.
}
\qianwen{}
\begin{itemize}[noitemsep, leftmargin=*,nosep]
    \item The first level, \textit{Task Planning}, serves as the strategic foundation where complex tasks are systematically decomposed into discrete, manageable sub-tasks. 
    \item The second level, \textit{Agent Assignment}, builds directly upon the task decomposition by assigning appropriate agents to each sub-task. 
    Some tasks may require a single agent, while others may necessitate collaboration between multiple agents with complementary roles. 
    \item The third level, \textit{Agent Optimization}, provides granular control mechanisms, allowing users to fine-tune individual agents through prompt engineering, tool integration, or access to external datasets, to precisely shape their behavior and response characteristics. 
\end{itemize}
These hierarchical levels reflect a natural design progression, from abstract strategy to detailed implementation, and form the structural backbone of the visual exploration offered by \sysName{}

\begin{figure*}
    \centering
    \includegraphics[width=0.95\linewidth]{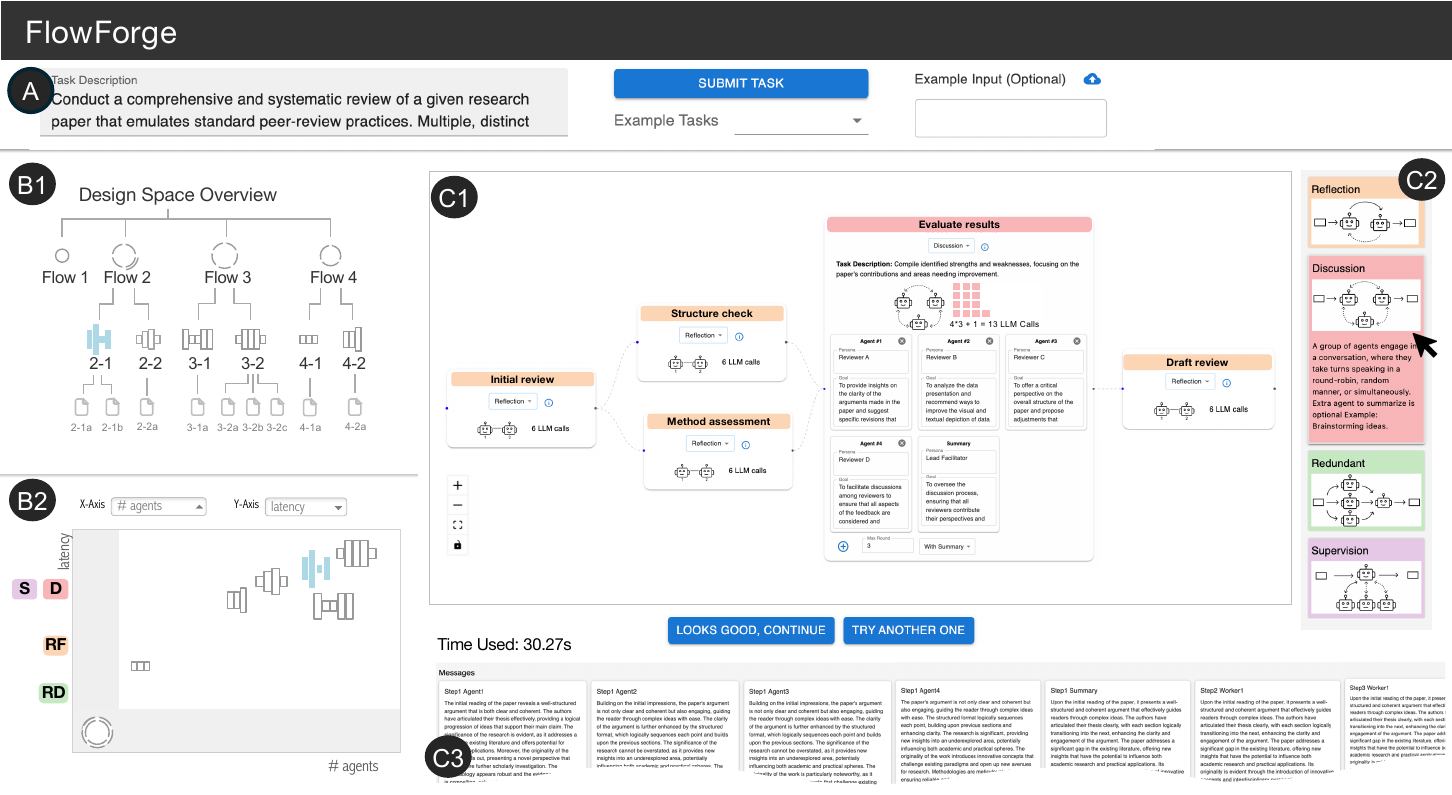}
    \vspace{-1em}
    \caption{\textbf{Interface Overview}. Users start by entering a task description (A). \sysName{} supports workflow creation through two main views: the \textit{Design Space View}, with a hierarchical tree (B1) and a scatter plot (B2), and the \textit{Canvas View}, which shows workflow details (C1), in-situ design suggestions (C2), and execution results (C3). }
    \label{fig:interface}
\end{figure*}

\subsection{Design Goals}
\sysName{} is developed with the following design goals in mind, informed directly by the challenges uncovered in our formative study: 
\begin{enumerate}
[leftmargin=*, label=\textbf{DG.\arabic*}., noitemsep, labelwidth=1em, ]
    \item \textbf{Step-by-step guidance from abstract to concrete.} 
    \label{DG:step-by-step}
    \sysName{} supports a top-down design process by structuring the workflow creation experience, guiding users from high-level task decomposition to granular agent optimization.
    This helps users navigate the complexity of the design space and mitigates unstructured exploration (\ref{challenge:vast}, \ref{challenge:unstructured}). 
    \item \textbf{Support for divergent and convergent thinking.}
    \label{DG:divergent}
    The system encourages users to explore multiple workflow alternatives early in the design process before committing to a final solution. 
    This prevents them from design fixation and promotes more deliberate decision-making (\ref{challenge:fixation}).
    \item \textbf{In-situ guidance for best practices.} 
    \label{DG:in-situ}
    As users explore the design space, \sysName{} provides context-aware design pattern suggestions drawn from validated design patterns.
    These suggestions surfaced at appropriate abstraction levels, assisting users without disrupting their flow (\ref{challenge:vast}\ref{challenge:unstructured}).
    \item \textbf{Parallel exploration across multiple dimensions}.
    \label{DG:compare}
    Users can evaluate and compare different workflows across key performance dimensions (\eg, latency, accuracy, cost) as well as task-specific dimensions (\eg, format constraints), enabling a more holistic understanding of trade-offs and priorities (\ref{challenge:metrics}). 
 \end{enumerate}

\section{System}


\sysName{} is an interactive visualization system that supports the design and exploration of multi-agent workflows. 
As shown in \cref{fig:interface}, users begin by entering a task description (A), after which \sysName{} generates and visualizes candidate workflows. The interface is composed of two main components: the \textit{Design Space View}, which displays the levels and dimensions via a hierarchical tree (B1) and a scatter plot (B2); and the \textit{Canvas View}, which presents the detailed structure of a selected workflow (C1), along with in-situ design suggestions (C2) and execution results (C3).

\subsection{Design Space View: Visualizing Abstraction Levels}
The hierarchical tree (\cref{fig:interface}.B1) captures three levels of abstraction in workflow design. 
This tree dynamically updates as users navigate the space, guiding them through the progressive stages of workflow design: task planning, agent assignment, and agent optimization (\ref{DG:step-by-step}). 
This structure encourages both depth-first exploration (\ie, jump directly to concrete implementation) and breath-first exploration (\ie, comparing options before committing), as the latter is often overlooked in current practices (\ref{DG:divergent}).
For instance, at the task planning level, users may compare task decompositions with three vs. four subtasks. At the agent assignment level, they can explore collaboration strategies, such as having two agents co-work on a subtask versus one completing it and another reviewing the output.

\begin{figure}
    \centering
    \includegraphics[width=0.9\linewidth]{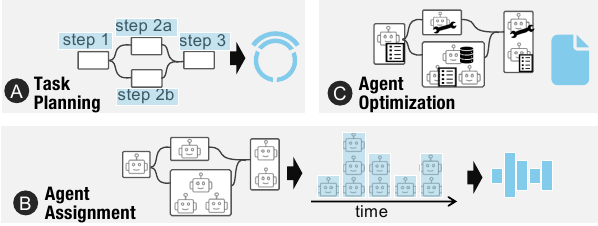}
    \caption{\textbf{Glyphs}. (A) At Level 1 (task planning), arcs represent subtasks and node size for sequential count. (B) At Level 2 (agent assignment), bars show agent calls over pseudo-time steps. (C) At Level 3 (agent optimization), workflows are denoted as file icons, as no new structural information is added and the workflow is fully runnable.  }
    \label{fig:glyph}
\end{figure}

To represent workflows across these levels, we introduce a set of visual glyph nodes (\cref{fig:glyph}).
These glyph nodes encode the \textit{computational efficiency}, a universally relevant dimension in workflow design (Sect.\ref{subsubsec:dimensions}) that can be estimated before execution.
\begin{itemize}[noitemsep, leftmargin=*,nosep]
    \item Level 1 (Task Planning): Glyphs use the number of arcs to indicate the number of subtasks. Stacked arcs represent parallel subtasks, and node size reflects the number of sequential steps.
    \item Level 2 (Agent Assignment): Glyphs consist of bars, where each bar represents a pseudo-time step of the workflow running process, and its height indicates the number of concurrent agent calls.
    \item Level 3 (Agent Optimization): As this level does not introduce structural changes, glyphs are rendered as file icons, denoting a fully runnable workflow.
\end{itemize}
These glyphs provide a quick, visual summary of each workflow, helping users efficiently compare alternatives without delving into detailed specifications.
While we experimented with encoding additional dimensions into the glyphs, we found that increasing complexity introduced unnecessary cognitive burdens.



\subsection{Design Space View: Visualizing Design Dimensions}
Although the hierarchical tree organizes workflows by abstraction level, it offers limited visibility into design dimensions (\eg, number of agents, number of subtasks, number of LLM calls, running time).
To address this, \sysName{} includes a configurable scatter plot (\cref{fig:interface}.B2) that enables comparison of workflows across key dimensions (\ref{DG:compare}).
In this scatter plot, users can assign different dimensions to the x- and y-axes.
We provide a list of predefined dimensions based on feedback from formative studies, while allowing users to add customized dimensions (\eg, user rating for the workflow outputs) via manually annotating the workflow outputs.
Workflows with undefined values on a selected axis are displayed in gray margin areas. 
For example, in \cref{fig:teaser}. A2, workflows at Level 1 (task planning) do not include information about the number of agents (y-axis), and are put in the bottom gray area of the scatter plot.
The scatter plot only displays workflows at the current level or one level above the selected node in the tree. 
This selective display is informed by the user feedback we received during the iterative design process, as showing all workflows across three abstraction levels can be cognitively overwhelming.

\qianwen{
We chose a scatter plot because it provides a clear overview of the design space, making it easier to identify regions that are heavily explored (\ie, clusters) versus those that remain underexplored (\ie, spatial gaps). 
For example, \cref{fig:interface} shows that the current exploration has exclusively focused on workflows that have multiple agents, as evidenced by the glyph nodes clustered in the upper-right corner. 
This insight may prompt users to consider exploring workflows with fewer subtasks or simpler configurations that have been overlooked (\ref{DG:divergent}).
}

\qianwen{We have explored alternative designs such as tables, parallel coordinates, and connected scatter plots. 
Although tables and parallel coordinates can display more than two dimensions simultaneously, users often struggle to interpret and compare more than two dimensions effectively \cite{wang2019atmseer}.
During the iterative design process, we also experimented explicitly connecting related workflows across levels in the scatter plot.
However, the additional edges in the scatter plot were unnecessary in most cases, since the children node typically share some dimension values as their parent nodes and their hierarchical relation can be easily recognized based on the alignment of the glyph nodes, \eg, same x position in the \cref{fig:teaser}.A2. 
}

\begin{figure}
    \centering
    \includegraphics[width=\linewidth]{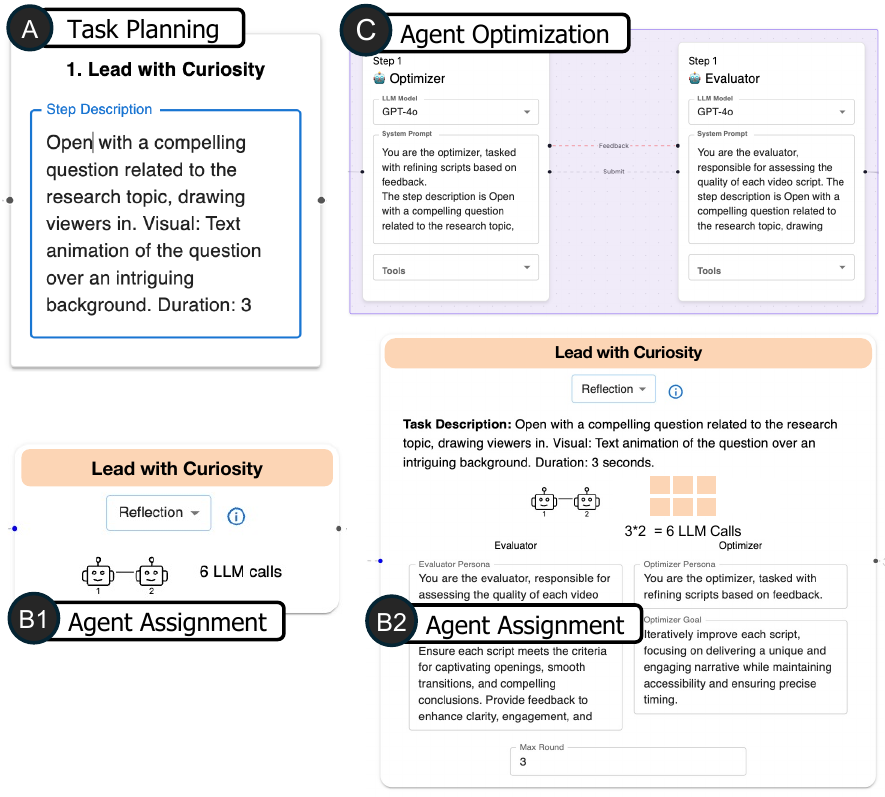}
    \caption{In the \textit{Canvas View}, the visual representation of each node adapts based on the abstraction level.}
    \label{fig:canvas}
\end{figure}

\subsection{Canvas View: Editing Individual Workflows}
Workflows selected from the design space are displayed in the \textit{Canvas View}, as interactive node-link diagrams.
Users can navigate between abstraction levels (\ref{DG:step-by-step}), using the \textit{``Looks Good, Continue''} and \textit{``Try Another One''} buttons in the \textit{Canvas View}.
The former advances the current workflow to a more concrete level (\eg, from task planning to agent assignment), while the latter prompts alternative solutions at the same level (\eg, a different way to decompose the task).
To prevent potential confusion between the tree representation of the design space and the node-link diagram of an individual workflow, different visualizations are used: circular nodes and step-wise edges for the tree and rectangular nodes with curved edges for the workflow.
Node styles of the workflow adapt according to the abstraction level selected in the left-side \textit{Design Space View}:
\begin{itemize}[noitemsep, leftmargin=*,nosep]
    \item Level 1 (Task Planning): Nodes represent subtasks with editable text descriptions. Edges denote dependencies and execution order of subtasks, with the flow progressing from left to right. Users can modify task structure, text, and dependencies.
    \item Level 2 (Agent Assignment): Nodes correspond to agent groups assigned to subtasks. When zoomed out (B1), nodes show summary information, \ie, the subtask name, the associated collaboration pattern, and the estimated number of LLM calls. Zooming in (B2) reveals detailed configurations, such as agent roles, collaboration patterns (\eg, reflection, supervision) (\autoref{tab:design_patterns}), and pattern-specific parameters (\eg, number of agents in redundancy or discussion rounds).
    \item Level 3 (Agent Optimization): Nodes represent individual agents, and agents working on the same subtask are grouped within the same purple rectangle. Users can configure agents using common patterns, including prompt engineering, tool usage, and retrieval augmentation.
\end{itemize}


\subsection{Design Cards: In-situ Guidance via Design Patterns}

To support in-situ design thinking, \sysName{} surfaces relevant design patterns on the right side of the \textit{Canvas View}.
Each pattern is presented as a card with its name, visual representation, and expandable details (\cref{fig:interface}.C2). 
When users hover over a pattern card, all corresponding nodes adopting the pattern are highlighted. 


Since different design patterns tend to prioritize different performance dimensions, we also annotate the scatter plot axes with relevant patterns when one of these dimensions is used as an axis. 
For example, as shown in \cref{fig:interface}.B2, when latency is selected as the y-axis, patterns such as Redundancy, Reflection, and Discussion, are recommended and marked along the axis to aid interpretation: Redundancy at the bottom (minimal latency), Reflection mid-range (moderate latency), and Discussion near the top (high latency due to multi-round interactions).

\subsection{Workflow and Suggestion Generation}
\qianwen{
The generation approach employs a straightforward, prompt-based methodology using GPT-4o~\cite{openai2024gpt4ocard}, as
existing methods are either not open-source or limited to specific tasks~\cite{hong2024metagpt}.
This prompt-based approach proves sufficient for demonstrating the effectiveness of \sysName{}. 
While we acknowledged that the generation can be further improved with more advanced approaches (\eg, reinforcement learning), novel methods for automatic workflow generation and recommendation is beyond the scope of this study.
}


\qianwen{Specifically, \sysName{} generates workflows and design patterns following the three levels of abstraction:}
\begin{itemize}[noitemsep, leftmargin=*,nosep]
    \item \qianwen{Level 1 (Task Planning): 
    \sysName{} prompts GPT-4o to generate diverse decompositions of a given task while maintaining a unified format.
    Each generated subtask is formatted to include a label, task description, output format, and a pointer to the next step.
    The prompt encourages variation in complexity (number of subtasks), structure (sequential or parallel), and perspective (distinct focus for each plan).
    Users can customize the decomposition by editing, adding, or deleting subtasks via node modifications, and reorder them by reconnecting edges in the Canvas View.
    }
    \item  \qianwen{Level 2 (Agent Assignment): 
    GPT-4o is prompted to rank a given list of design patterns for each subtask and provide explanations of its recommendation. 
    For each subtask, the prompt combines its context (task name and description) with the full list of design patterns (names, definitions, and brief examples). 
    Based on this information, GPT-4o heuristically evaluates the subtask's characteristics and suggests suitable patterns,
    \eg, Redundant for multifaceted analysis.
    The combination of design patterns is randomized to mitigate the position bias of LLM~\cite{zheng2023judging}. 
    Users can view the description and estimated computational cost of the suggested patterns, modify the design pattern for each subtask (\eg, change from Discussion to Reflection), and customize its corresponding parameters (\eg, number of agents).}
    \item \qianwen{Level 3 (Agent Optimization): 
    GPT-4o is prompted to generate configuration details of each agent, including persona, goals, input and output format, based on the subtask's input and output, the selected design pattern, as well as available tools and external databases.
    Multiple agents within one pattern are encouraged to embody distinct perspectives that align with real-world roles.
    Users may refine prompt content, the underlying LLM, the tool usage, and the external database of a specific agent.
    }
\end{itemize}
Full implementation details are available in the project website.

\subsection{Implementation}
\sysName{} is implemented in JavaScript using React for building the user interface, D3~\cite{d3} for rendering custom data-driven visualizations, and ReactFlow~\cite{reactflow} for managing and displaying the individual multi-agent workflows as interactive node-link diagrams. 
We use LangGraph \cite{langgraph} to store and execute the generated workflows. 
\sysName{} maintains its own internal data structure and uses a separate converter to translate this structure into compatible LangGraph files. 
This modular design allows \sysName{} to be easily adapted to other workflow frameworks by modifying the converter to support other formats.
The source code, online demo, and documentation of \sysName{} are available at 
\url{https://visual-intelligence-umn.github.io/FlowForge}.
\section{Case Studies}
To demonstrate the utility of \sysName{} in real-world scenarios, we present two example use cases of \sysName{}, developed in collaboration with LLM researchers (Section.\ref{subsec:understanding}). 

\begin{figure}
    \centering
    \includegraphics[width=\linewidth]{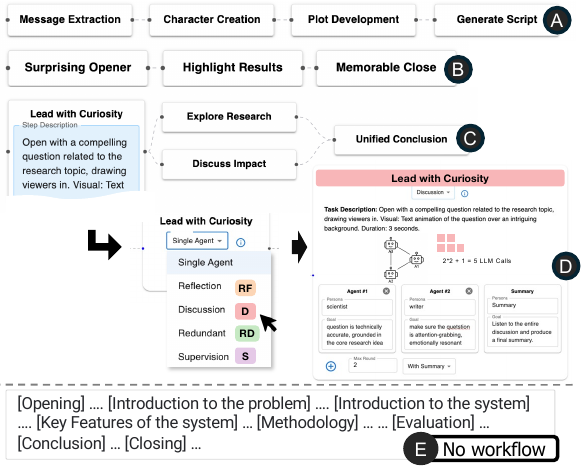}
    \caption{\textbf{Case Study 1: Fast-Forward Video Planning.} (A-D) \sysName{} helps users create multiple creative ways of generating the transcript of fast-forward videos for a research paper. (E) In contrast, directly prompting GPT-4o often produces transcripts with conventional structures and exceeds the 25-second constraints. Detailed text responses are omitted due to space constraints.}
    \label{fig:case-video}
\end{figure}

\subsection{Fast-Forward Video Generation for Research Papers}
\label{subsec:case-video}
In this use case, a user wants to create a multi-agent workflow for generating ideas for a 25-second fast-forward video that advertises a research paper. This case demonstrates \sysName{}'s ability to improve diversity and creativity of workflow designs.

After submitting a task description, \sysName{} generates five different task planning options with varying lengths.
%
While these workflows follow a similar structure comprising a beginning, an optional middle section, and an ending, 
they differ in how each subtask is defined and executed.
For example, some workflows begin with an intriguing question (\cref{fig:case-video}.C), while others open with a surprising research finding (\cref{fig:case-video}.B). 
One particularly creative workflow introduces a storytelling element to engage the audience (\cref{fig:case-video}.A). It begins with a ``message extraction'' step to identify the paper’s core contribution, followed by three narrative-oriented subtasks: character creation, plot development, and script preparation.
Ultimately, the user selects a workflow that incorporates a parallel structure in the middle section (\cref{fig:case-video}.C). This design enables two alternative narrative paths: one that explains the research method and another that emphasizes the broader impact of the work. The workflow can then evaluate the outputs of both paths and select the more compelling version for the final fast-forward video.

With the task planning finalized, the user proceeds to the agent assignment level. Most subtasks are assigned to single agents, likely because they are simple and well-defined. 
However, for crafting the video’s opening, where creativity is key, the user wants to encourage a more dynamic and collaborative process. 
To inform this decision, the user configures the scatter plot to display ``Creativity'' along the y-axis.
As revealed in the scatter plot, certain design patterns, such as Discussion and Redundancy, are associated with higher creativity scores.
Based on this, the user replaces the single-agent setup with a Discussion pattern, enabling multiple agents to brainstorm and refine the opening collaboratively (\cref{fig:case-video}.D).
In the final stage, the user enters the agent optimization level and fine-tunes the prompts for each agent to better control the expected outputs. The workflow is then executed within \sysName{}, where both the runtime and outputs are automatically recorded. Users can rate the results, iterate by modifying agent prompts, and compare different versions in the scatter plot across different dimensions.

\noindent
\textbf{Comparison with a one-pass LLM.} 
While a single GPT-4o prompt can generate a video script, it typically defaults to conventional structures (\eg, introduction, method, conclusion), and often fails to respect format constraints such as time limits. In our test, even with clear instructions for a 25-second limit, GPT-4o generated long-form outputs mimicking full paper summaries with five or more sections (\cref{fig:case-video}.E).
In contrast, \sysName{} empowers users to explore varied narrative structures (\cref{fig:case-video}.A-C), experiment with different decompositions, and fine-tune agent behaviors, offering greater control, diversity, and creativity in workflows generation.

\subsection{Data Storytelling from a Dataset}
In this case study, a user creates a workflow to generate visualizations and produce data-driven storytelling based on a given dataset. We use a Movies dataset~\cite{vegaLiteMovies} as the input to evaluate the outputs of different workflow configurations. 
This case demonstrates \sysName{}’s ability to compare different workflows and their generated outputs.

\begin{figure}
    \centering
    \includegraphics[width=\linewidth]{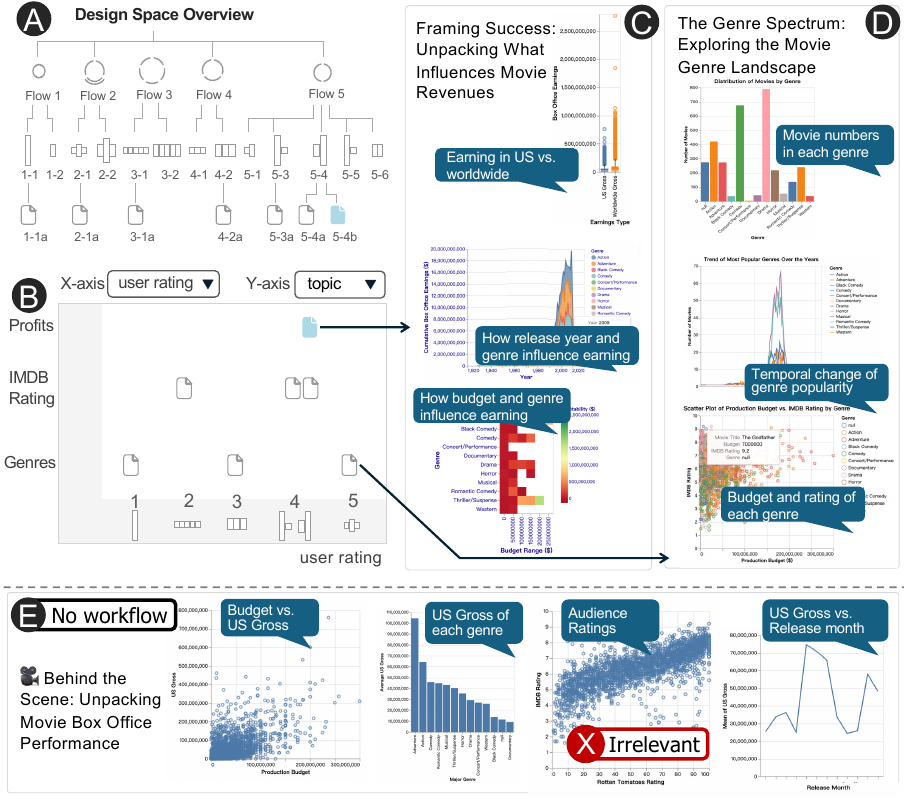}
    \caption{\textbf{Case Study 2: Visualization Generation}. (A–B) \sysName{} enables users to explore and compare different workflows for generating data-driven visual stories from a given dataset. (C–D) Example outputs from these workflows. (E) Directly prompting GPT-4o generates visualizations irrelevant to the data story topic. Due to space constraints, detailed text narration for each visualization is not included.}
    \label{fig:case-vis}
    \vspace{-1em}
\end{figure}

As shown in \cref{fig:case-vis}.A, the user explores a variety of task planning and agent configurations. 
One specific task planning option, Flow 5, is explored more extensively, indicated by its five child nodes in the hierarchical tree view. 
This workflow contains only two subtasks (represented by two arcs in the glyph), which the user favors due to lower latency. 
At the agent assignment level, the user evaluates multiple alternatives and selects only a subset to refine further at the agent optimization level, which is reflected by the fact that only certain glyphs have child nodes in the hierarchical tree.
When the workflow has fewer subtasks, the user tends to prefer configurations with more agents per step (indicated by higher bars), such as choosing Flow 1-1 over 1-2 in \cref{fig:case-vis}.A. 
Conversely, for workflows with more subtasks, the user tends to favor configurations with fewer agents per step (indicated by shorter bars), such as selecting Flow 3-1 over 3-2.

To arrange and compare different workflows, the user configures the x- and y-axes of the scatter plot (\cref{fig:case-vis}B). These axes can represent system-level metrics such as runtime and token usage, or custom metrics defined by the user, such as subjective ratings of output quality or the topic of the generated storytelling.

As shown in \cref{fig:case-vis}B, the workflow outputs vary in both topic and quality. Glyphs at Level 2 (agent assignment) appear in the bottom gray area of the scatter plot, as they lack the topic attribute required for comparison. Their horizontal positions reflect the average user rating of each workflow. 
Interestingly, in this experiment, neither increasing the number of steps nor adding more agents consistently led to better storytelling outcomes. \cref{fig:case-vis}.C–D present example outputs from two selected workflows, illustrating the variability in topic focus.

\noindent
\textbf{Comparing with one-pass LLM.} 
When prompting a single LLM (GPT-4o), the model often failed to strictly follow instructions: create visualizations centered on a specific topic with a coherent narrative. 
For example, in our experiments, GPT-4o generated a data-driven story titled \textit{``Behind the Scene: Unpacking Movie Box Office Performance''}.
However, the outputs included unrelated visualizations that were not aligned with the intended focus, such as a scatter plot comparing Rotten Tomatoes Rating against IMDB Rating for each movie (\cref{fig:case-vis}.E).
In contrast, \sysName{} supports focused, modular task decomposition, enabling users to guide narrative flow, specify visualization goals, and experiment with structure, all of which contribute to higher coherence and alignment with user intent.

\section{User Studies}

To evaluate how \sysName{} supports the design experience of creating multi-agent workflows, we conducted an observational study comparing it against LangGraph Studio~\cite{langgraphstudio2024}, a widely-used state-of-the-art interface for creating multi-agent workflows.

\subsection{Study Design}

\noindent
\textbf{Participants}.
We recruited nine participants (seven male, two female; seven aged 18–24, two aged 25–34) via email list and personal contact.
All participants have prior knowledge about multi-agent LLM workflows but had no prior exposure to this project.
Each study session lasted approximately one hour, and participants received a \$10 Amazon gift card as compensation. 
The study was approved by the Institutional Review Board, and all participants signed an informed consent form. 

\begin{figure}
    \centering
    \includegraphics[width=\linewidth]{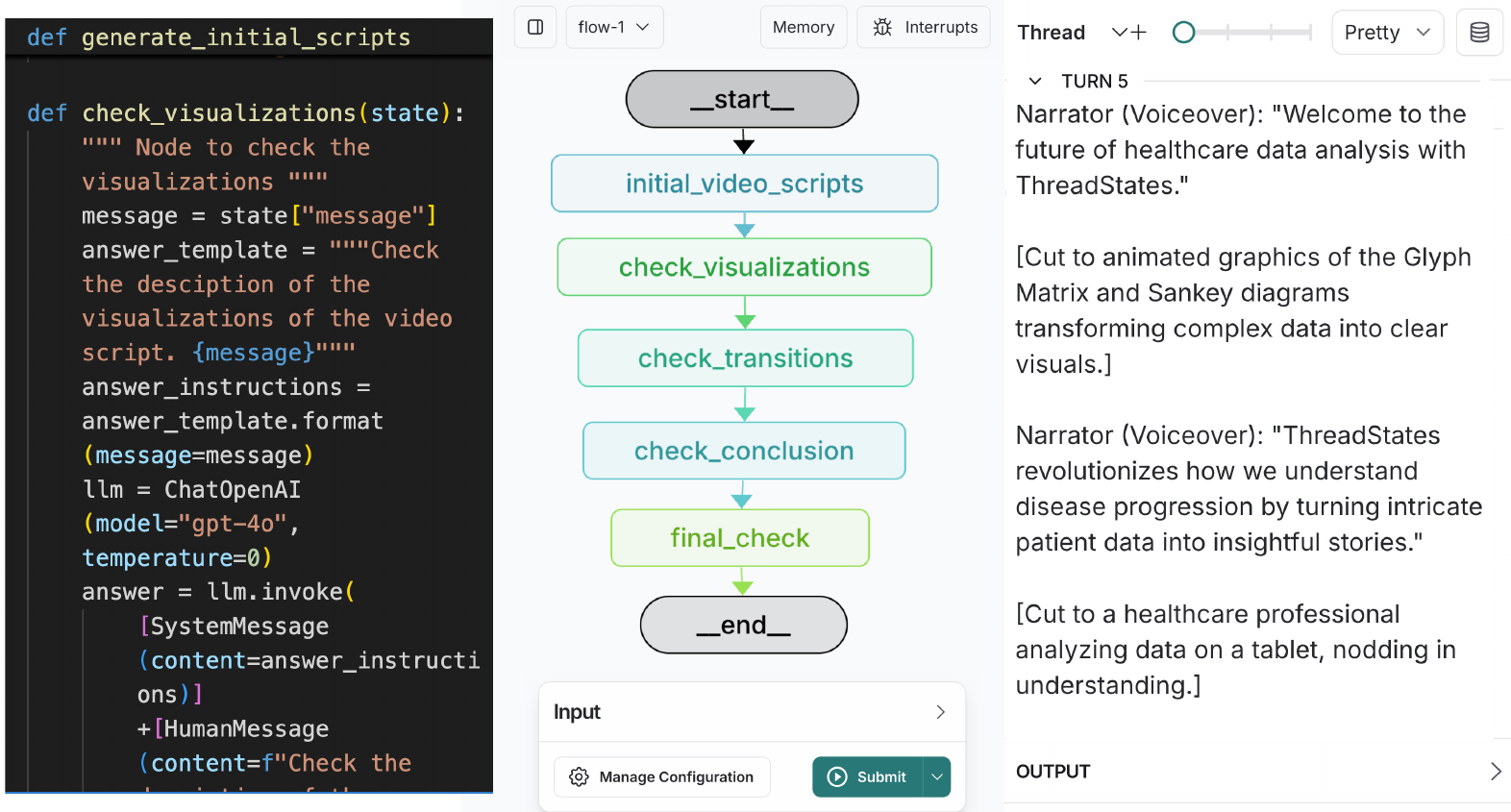}
    \caption{\textbf{User Study Baseline}. We use LangGraph Studio as the baseline system in the user study. 
    }
    \label{fig:baseline}
\end{figure}

\noindent
\textbf{Baseline}.
For the baseline condition, we used LangGraph Studio (\cref{fig:baseline}), a visual programming tool for building and executing multi-agent workflows.
While highly expressive and widely adapted, LangGraph Studio does not provide structured design scaffolding or exploratory support, which are key features of \sysName{}. 


\begin{figure}
    \centering
    \includegraphics[width=\linewidth]{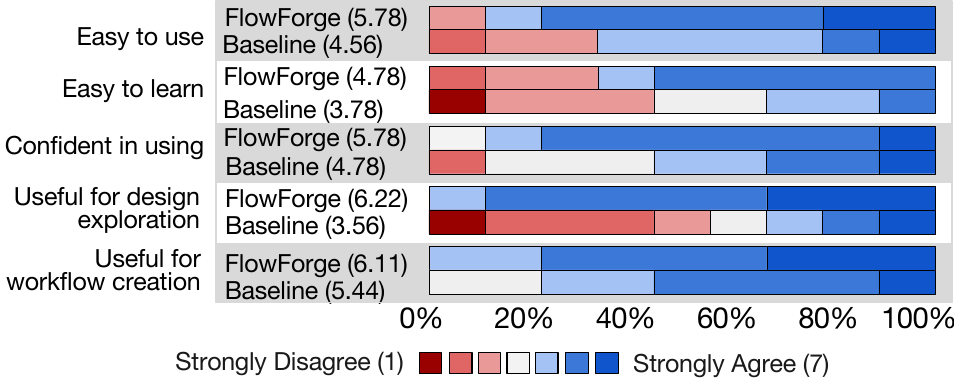}
    \caption{\textbf{User Ratings}. Comparison of \sysName{} and the baseline system based on post-study questionnaire responses. Participants rated each item on a 7-point Likert scale ranging from strongly disagree (1) to strongly agree (7). Numbers in parentheses indicate the average rating.}
    \label{fig:result}
\end{figure}

\noindent
\qianwen{
\textbf{Task \& Study Procedure}.
In both conditions (\sysName{} vs. baseline), participants created workflows for the same task: generating scripts for a 25-second fast-forward video that introduces a research paper. 
Prior to the study, participants received comprehensive orientation including task explanation, example papers for workflow evaluation, and detailed prompting specifications describing the desired outputs.
During the study, participants completed both conditions in randomized order to control for potential learning effects.
Targeted tutorials and demonstrations for the corresponding system are provided before beginning each condition.
In both conditions, participants constructed workflows from scratch without any default templates, with explicit encouragement to create and compare multiple workflows. 
The session of a condition ended when participants expressed satisfaction with their results or upon reaching the 25-minute time limit, whichever occurred first.
Participants were instructed to think aloud throughout the sessions and can ask questions freely at any time.
After two conditions, participants completed a usability questionnaire and a post-study interview.
}

\noindent
\textbf{Data Analysis.}
We analyzed (i) the workflows created by participants, including the time participants spent creating them and their overall quality; (ii) self-reported usability ratings for both systems; and (iii) participants’ think-aloud protocols, interactions during the study, and qualitative feedback from post-study interviews.

\subsection{Quantitative Results}
Our quantitative evaluations demonstrated that \sysName{} outperformed the baseline system, delivering improvements in both user experience and workflow creation efficiency.

First, \sysName{} led to improved user experience in creating LLM workflows.
\cref{fig:result} presents participants’ subjective ratings comparing \sysName{} and the baseline across five dimensions related to usability and utility. 
Overall, \sysName{} was rated more favorably. Participants found it easier to use, with a higher proportion agreeing or strongly agreeing that they would not require technical support to operate the system. They also reported greater confidence when using \sysName{}, as well as stronger agreement on its usefulness for both exploring alternative workflows and executing them.

\begin{figure}
    \centering
    \includegraphics[width=\linewidth]{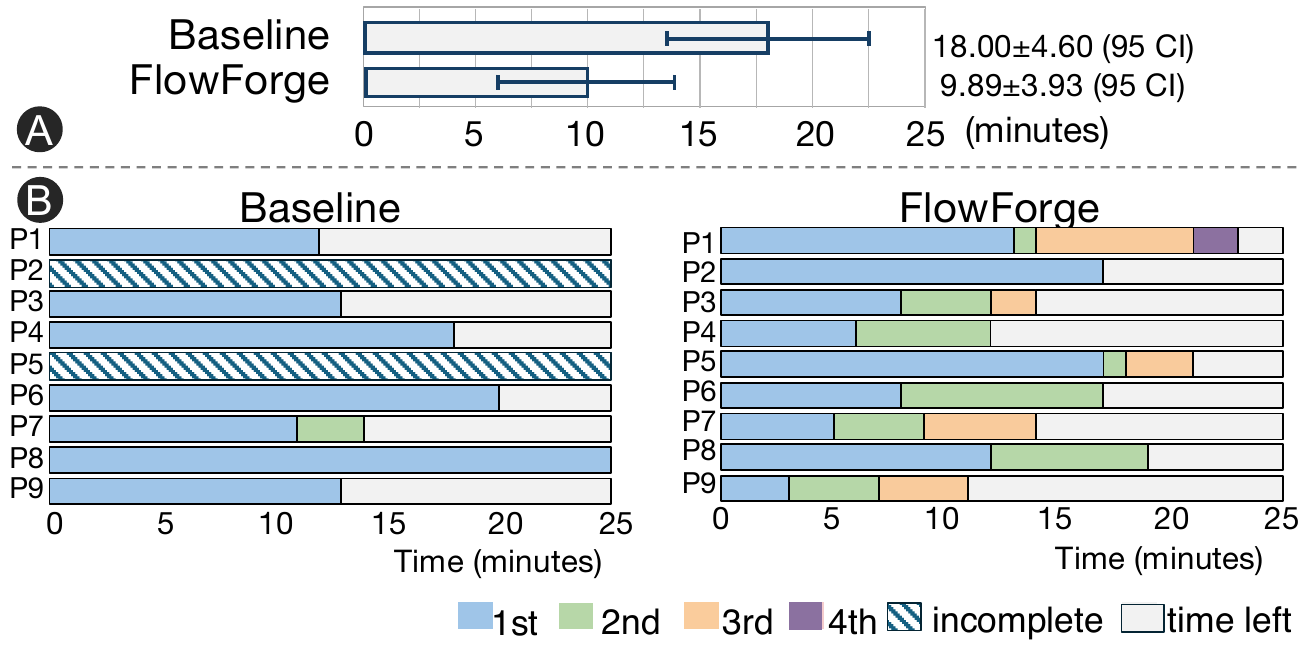}
    \caption{\textbf{Workflow Creation Time and Quantity}. (A) Users spent less time creating their first runnable workflow in \sysName{} compared to the baseline. Error bars indicate 95\% confidence intervals (CI). (B) Users created more workflows in \sysName{} compared with the baseline.}
    \label{fig:time}
\end{figure}

Second, \sysName{} enabled more efficient workflow creation with more diverse designs.
\cref{fig:time} presents participants' workflow creation time. 
Participants spent on average 9.89 (95\% CI: 5.96-13.82) minutes to create and run the first workflow in \sysName{}, and 18.0 (95\% CI: 13.40-22.60) minutes in the baseline. 
Student's t-test showed a significant difference between their completion time for the first workflow (\textit{t}=-8.1, \textit{p}=.003). Most (8/9) of the participants were able to create and explore more than one workflow in \sysName{} after the initial try, while some participants (2) failed to create a workflow in the baseline. 
For participants who did not finish, 25 minutes (maximum allowed time per session) was used as their estimated completion time.


To evaluate workflow diversity, we conducted a manual analysis of workflows created by participants across both conditions. 
While workflows created in both conditions contained a similar number of subtasks (baseline: 3.43, \sysName{}: 3.89), \sysName{} produced more parallel subtasks (3) than the baseline (1).
Participants also created more diverse agent communications using \sysName{}. Using \sysName{}, participants implemented a variety of agent communication patterns (7 reflection, 5 supervision, 7 redundancy, 8 discussion, out of 35 total instances). In contrast, baseline workflows exclusively employed single-agent across all 22 instances.




\subsection{Observations and Feedback}

Based on the think-aloud comments and interviews, we reported how participants approached workflow creation across the two conditions.

Across both conditions, participants prioritized creating an initial functional workflow before exploring alternatives. 
Participants appreciated the guidance provided at each level. 
They noted that while Levels 2 and 3 of the design guidance were more \textit{``technically oriented''}, Level 1 introduced innovative concepts that were accessible and valuable even without specialized AI knowledge.

\qianwen{Participants demonstrated iterative workflow design behaviors in both conditions, yet their approaches differed substantially.
In the baseline condition, 
participants typically constructed workflows with only one or two sequential subtasks.
When refining these workflows, they focused exclusively on prompt refinement for individual agents while maintaining the original workflow structure unchanged. 
When using \sysName{}, participants exhibited broader explorations. 
Most participants (8/9) explored different task planning (level 1) to generate alternative workflows and compare workflows based on key dimensions such as output quality and latency. 
Two participants made strategic modifications when encountering performance issues. 
Informed by the design cards at level 3,
P5 improved output quality by incorporating detailed domain expertise into agent personas. 
Through analysis of latency of different design patterns in the \textit{Design Space View},
P9 transitioned from the Discussion pattern to Redundant and restructured task planning to enable parallel subtask execution, successfully reducing execution time while preserving output quality.
}

\section{Discussion}

\subsection{Design Implications.}
We reflect on the design implications of \sysName{} by connecting it with previous visualization tools developed for AI research.

\noindent
\textbf{Design Space Visualization.}
This work presents an initial exploration of using design space visualization to support the navigation and construction of complex multi-agent workflows. 
Although visualization of the design space has been shown as an effective approach in AI applications such as deep neural network design~\cite{wang2019visual}, automated machine learning~\cite{wang2019atmseer}, and neural architecture search~\cite{yuan2022visual}, it has not yet been applied to the emerging context of multi-agent LLM workflows.
This new context presents unique opportunities. 
Compared to traditional AI development, which often requires deep technical expertise to understand the design space, LLM workflows significantly lower the barrier to entry, allowing a broader range of end users to design AI-driven systems. 
As this shift moves design tools from expert-centric to end-user-facing contexts, visualization must evolve to offer greater clarity and interpretability.

\noindent
\textbf{Thinking Scaffold for both LLMs and humans.}
While \sysName{} is designed for building multi-agent workflows, we observed that human users also benefit from the thinking scaffold it provides, particularly in open-ended and creative tasks (\eg,  creatively crafting fast-forward videos using character and plot elements in \cref{subsec:case-video}). 
By explicitly visualizing workflows, users engage in cognitive externalization that aligns with extended mind theory, where cognitive processes are distributed across both internal mental representations and external artifacts, making it easier to brainstorm effectively with LLMs. 

This approach represents a different perspective from traditional visualization for AI interpretability, which has focused on visualizing low-level computational processes (\eg, neuron activations) \cite{jin2022gnnlens,liu2016towards, lu2024gnn}. 
The emerging of LLMs has enabled new interpretability opportunities by exporting reasoning in workflows and agent interactions \cite{wu2022chains}. Interactive visualizations of these reasoning processes can thus serve as accessible collaborative interfaces moving beyond expert-only tools.


\subsection{Limitations and Future Directions.}

\qianwen{First, \sysName{} does not cover the full spectrum of workflow designs and application scenarios. 
Regarding workflow architecture, \sysName{} focuses on workflows with static or conditional communication protocols but does not support dynamic architecture modifications during execution. 
For example, the system cannot support workflows that undergo real-time architecture revision based on human-in-the-loop interventions \cite{wang2024surveyagents} or agent feedback \cite{hu2024ADAS}. 
Regarding application scenarios, \sysName{} is specifically designed for structured problem-solving tasks rather than world simulation applications \cite{wang2024openworld, guo2024survey}, which may require fundamentally different architectural considerations and interaction patterns.
}


Second, in \sysName{}, we used straightforward prompting techniques for task decomposition and design pattern recommendation, as the primary focus of this study is to demonstrate the benefits of visualizing and interacting with the design space during workflow exploration. 
We acknowledge that the quality of generated workflows would be further improved with more advanced methods, such as using reinforcement learning to optimize the workflow design. 


\qianwen{
Finally, in spite of the intuitiveness and effectiveness, the three-level abstraction of design space is not the only viable scaffold and can be further enhanced.
For example, Liu \etal.\cite{liu2024cataloque} proposed a decision tree that covers key decisions for designing multi-agent LLM systems. Compared to these alternatives, our method aims to strike a balance between expressiveness (\ie, enabling diverse workflow constructions) and conciseness (\ie, avoiding excessive complexity).
There are several promising directions for further enhancing this abstraction. Each level may benefit from further decomposition into sublevels that capture more nuanced design choices. For example, the task planning level may involve sub-levels such as failure handling mechanism. Meanwhile, additional support can be provided to capture the inter-level dependencies and facilitate design iterations across abstraction levels~\cite{epperson2025interactive}, such as revisiting task planning decisions based on agent optimization.
}

\section{Conclusion}
This study presents \sysName{}, an interactive visualization tool that supports the creation of multi-agent LLM workflows by organizing the levels and dimensions of the design space, visualizing the design space and the various solutions within it, and integrating in-situ 
guidance from well-established design patterns in design exploration. 
Our evaluation demonstrated that \sysName{} improved user experience, increased workflow creation efficiency, and enhanced the diversity of created workflows. 
This work highlights the potential of design space visualization as a thinking scaffold for both LLMs and human designers in complex LLM system development.




\bibliographystyle{abbrv-doi-hyperref}

\bibliography{reference}

\begin{thebibliography}{10}

\bibitem{vellum}
{Cellum. Purpose-Built GUI and SDK for AI Development}.
\newblock \url{https://www.vellum.ai/}.

\bibitem{crewai}
{CrewAI: Build AI agent teams that work together to tackle complex tasks.}
\newblock \url{https://docs.crewai.com/introduction}.

\bibitem{rivet}
{Rivet. The Open-Source Visual AI Programming Environment}.
\newblock \url{https://rivet.ironcladapp.com/}.

\bibitem{aws2024orchestrator}
Amazon.
\newblock {Multi-Agent Orchestrator framework}.
\newblock \url{https://awslabs.github.io/multi-agent-orchestrator/}.

\bibitem{anthropic2024agents}
Anthropic.
\newblock {Building Effective Agents}.
\newblock \url{https://www.anthropic.com/research/building-effective-agents}, 2024.

\bibitem{d3}
M.~Bostock, V.~Ogievetsky, and J.~Heer.
\newblock {D3 Data-Driven Documents}.
\newblock {\em IEEE Transactions on Visualization and Computer Graphics}, 17(12):2301–2309, 2011. \href{https://doi.org/10.1109/TVCG.2011.185}
{doi: {{%
10\hspace{.1pt}\discretionary{.}{%
}{.}\hspace{.4pt}1109\discretionary{/}{%
}{/}TVCG\hspace{.1pt}\discretionary{.}{%
}{.}\hspace{.4pt}2011\hspace{.1pt}\discretionary{.}{%
}{.}\hspace{.4pt}185}}}


\bibitem{brade2023promptify}
S.~Brade, B.~Wang, M.~Sousa, S.~Oore, and T.~Grossman.
\newblock {Promptify: Text-to-Image Generation through Interactive Prompt Exploration with Large Language Models}.
\newblock In {\em Proceedings of the 36th Annual ACM Symposium on User Interface Software and Technology}, UIST '23,  article no. 96,  14 pages. ACM, New York, 2023. \href{https://doi.org/10.1145/3586183.3606725}
{doi: {{%
10\hspace{.1pt}\discretionary{.}{%
}{.}\hspace{.4pt}1145\discretionary{/}{%
}{/}3586183\hspace{.1pt}\discretionary{.}{%
}{.}\hspace{.4pt}3606725}}}


\bibitem{buxton2010sketching}
B.~Buxton.
\newblock {\em {Sketching user experiences: getting the design right and the right design}}.
\newblock Morgan Kaufmann, San Francisco, 2010. \href{https://doi.org/10.1016/B978-0-12-374037-3.X5043-3}
{doi: {{%
10\hspace{.1pt}\discretionary{.}{%
}{.}\hspace{.4pt}1016\discretionary{/}{%
}{/}B978\discretionary{%
}{-}{-}0\discretionary{%
}{-}{-}12\discretionary{%
}{-}{-}374037\discretionary{%
}{-}{-}3\hspace{.1pt}\discretionary{.}{%
}{.}\hspace{.4pt}X5043\discretionary{%
}{-}{-}3}}}


\bibitem{cemri2025multi}
M.~Cemri, M.~Z. Pan, S.~Yang, L.~A. Agrawal, B.~Chopra, R.~Tiwari, K.~Keutzer, A.~Parameswaran, D.~Klein, K.~Ramchandran, M.~Zaharia, J.~E. Gonzalez, and I.~Stoica.
\newblock {Why Do Multi-Agent LLM Systems Fail?}, 2025. \href{https://doi.org/10.48550/arXiv.2503.13657}
{doi: {{%
10\hspace{.1pt}\discretionary{.}{%
}{.}\hspace{.4pt}48550\discretionary{/}{%
}{/}arXiv\hspace{.1pt}\discretionary{.}{%
}{.}\hspace{.4pt}2503\hspace{.1pt}\discretionary{.}{%
}{.}\hspace{.4pt}13657}}}


\bibitem{chan2023chateval}
C.-M. Chan, W.~Chen, Y.~Su, J.~Yu, W.~Xue, S.~Zhang, J.~Fu, and Z.~Liu.
\newblock {ChatEval: Towards Better LLM-based Evaluators through Multi-Agent Debate}, 2023. \href{https://doi.org/10.48550/arXiv.2308.07201}
{doi: {{%
10\hspace{.1pt}\discretionary{.}{%
}{.}\hspace{.4pt}48550\discretionary{/}{%
}{/}arXiv\hspace{.1pt}\discretionary{.}{%
}{.}\hspace{.4pt}2308\hspace{.1pt}\discretionary{.}{%
}{.}\hspace{.4pt}07201}}}


\bibitem{dibia2024autogenstuido}
V.~Dibia, J.~Chen, G.~Bansal, S.~Syed, A.~Fourney, E.~Zhu, C.~Wang, and S.~Amershi.
\newblock {AUTOGEN STUDIO: A No-Code Developer Tool for Building and Debugging Multi-Agent Systems}.
\newblock In D.~I. Hernandez~Farias, T.~Hope, and M.~Li, eds., {\em Proceedings of the 2024 Conference on Empirical Methods in Natural Language Processing: System Demonstrations}, pp. 72--79. Association for Computational Linguistics, Miami, 2024. \href{https://doi.org/10.18653/v1/2024.emnlp-demo.8}
{doi: {{%
10\hspace{.1pt}\discretionary{.}{%
}{.}\hspace{.4pt}18653\discretionary{/}{%
}{/}v1\discretionary{/}{%
}{/}2024\hspace{.1pt}\discretionary{.}{%
}{.}\hspace{.4pt}emnlp\discretionary{%
}{-}{-}demo\hspace{.1pt}\discretionary{.}{%
}{.}\hspace{.4pt}8}}}


\bibitem{du2023mathdebate}
Y.~Du, S.~Li, A.~Torralba, J.~B. Tenenbaum, and I.~Mordatch.
\newblock {Improving Factuality and Reasoning in Language Models through Multiagent Debate}.
\newblock In {\em Proceedings of the 41st International Conference on Machine Learning}, ICML'24,  article no. 467,  31 pages. JMLR.org, 2024. \href{https://doi.org/10.5555/3692070.3692537}
{doi: {{%
10\hspace{.1pt}\discretionary{.}{%
}{.}\hspace{.4pt}5555\discretionary{/}{%
}{/}3692070\hspace{.1pt}\discretionary{.}{%
}{.}\hspace{.4pt}3692537}}}


\bibitem{ManusAI}
B.~Effect.
\newblock {Manus AI: A General AI Agent}.
\newblock \url{https://manus.im/}.
\newblock Accessed: 2025-03-18.

\bibitem{epperson2025interactive}
W.~Epperson, G.~Bansal, V.~C. Dibia, A.~Fourney, J.~Gerrits, E.~Zhu, and S.~Amershi.
\newblock Interactive debugging and steering of multi-agent ai systems.
\newblock In {\em Proceedings of the 2025 CHI Conference on Human Factors in Computing Systems}, pp. 1--15, 2025.

\bibitem{feng2023promptmagician}
Y.~Feng, X.~Wang, K.~K. Wong, S.~Wang, Y.~Lu, M.~Zhu, B.~Wang, and W.~Chen.
\newblock { PromptMagician: Interactive Prompt Engineering for Text-to-Image Creation }.
\newblock {\em IEEE Transactions on Visualization and Computer Graphics}, 30(01):295--305, 2024. \href{https://doi.org/10.1109/TVCG.2023.3327168}
{doi: {{%
10\hspace{.1pt}\discretionary{.}{%
}{.}\hspace{.4pt}1109\discretionary{/}{%
}{/}TVCG\hspace{.1pt}\discretionary{.}{%
}{.}\hspace{.4pt}2023\hspace{.1pt}\discretionary{.}{%
}{.}\hspace{.4pt}3327168}}}


\bibitem{reactflow}
W.~GmbH.
\newblock {React Flow: A library for building node-based UIs}.
\newblock \url{https://reactflow.dev/}, 2024.
\newblock Accessed: 2025-03-26.

\bibitem{gottweis2025towards}
J.~Gottweis, W.-H. Weng, A.~Daryin, T.~Tu, A.~Palepu, P.~Sirkovic, A.~Myaskovsky, F.~Weissenberger, K.~Rong, R.~Tanno, et~al.
\newblock {Towards an AI co-scientist}, 2025. \href{https://doi.org/10.48550/arXiv.2502.18864}
{doi: {{%
10\hspace{.1pt}\discretionary{.}{%
}{.}\hspace{.4pt}48550\discretionary{/}{%
}{/}arXiv\hspace{.1pt}\discretionary{.}{%
}{.}\hspace{.4pt}2502\hspace{.1pt}\discretionary{.}{%
}{.}\hspace{.4pt}18864}}}


\bibitem{Grunde-McLaughlin2024crowdsourcing}
M.~Grunde-McLaughlin, M.~S. Lam, R.~Krishna, D.~S. Weld, and J.~Heer.
\newblock {Designing LLM Chains by Adapting Techniques from Crowdsourcing Workflows}.
\newblock {\em ACM Transactions on Computer-Human Interaction}, 32(3),  article no. 27,  57 pages, 2025. \href{https://doi.org/10.1145/3716134}
{doi: {{%
10\hspace{.1pt}\discretionary{.}{%
}{.}\hspace{.4pt}1145\discretionary{/}{%
}{/}3716134}}}


\bibitem{guo2024survey}
T.~Guo, X.~Chen, Y.~Wang, R.~Chang, S.~Pei, N.~V. Chawla, O.~Wiest, and X.~Zhang.
\newblock {Large Language Model based Multi-Agents: A Survey of Progress and Challenges}, 2024. \href{https://doi.org/10.48550/arXiv.2402.01680}
{doi: {{%
10\hspace{.1pt}\discretionary{.}{%
}{.}\hspace{.4pt}48550\discretionary{/}{%
}{/}arXiv\hspace{.1pt}\discretionary{.}{%
}{.}\hspace{.4pt}2402\hspace{.1pt}\discretionary{.}{%
}{.}\hspace{.4pt}01680}}}


\bibitem{heidegger1977question}
M.~Heidegger.
\newblock {\em {The Question Concerning Technology, and Other Essays}}.
\newblock Harper \& Row, New York, 1977.

\bibitem{hong2024metagpt}
S.~Hong, M.~Zhuge, J.~Chen, X.~Zheng, Y.~Cheng, J.~Wang, C.~Zhang, Z.~Wang, S.~K.~S. Yau, Z.~Lin, L.~Zhou, C.~Ran, L.~Xiao, C.~Wu, and J.~Schmidhuber.
\newblock {MetaGPT: Meta Programming for A Multi-Agent Collaborative Framework}.
\newblock In {\em The Twelfth International Conference on Learning Representations}, 2024. \href{https://doi.org/10.48550/arXiv.2308.00352}
{doi: {{%
10\hspace{.1pt}\discretionary{.}{%
}{.}\hspace{.4pt}48550\discretionary{/}{%
}{/}arXiv\hspace{.1pt}\discretionary{.}{%
}{.}\hspace{.4pt}2308\hspace{.1pt}\discretionary{.}{%
}{.}\hspace{.4pt}00352}}}


\bibitem{hu2024ADAS}
S.~Hu, C.~Lu, and J.~Clune.
\newblock {Automated Design of Agentic Systems}, 2024. \href{https://doi.org/10.48550/arXiv.2408.08435}
{doi: {{%
10\hspace{.1pt}\discretionary{.}{%
}{.}\hspace{.4pt}48550\discretionary{/}{%
}{/}arXiv\hspace{.1pt}\discretionary{.}{%
}{.}\hspace{.4pt}2408\hspace{.1pt}\discretionary{.}{%
}{.}\hspace{.4pt}08435}}}


\bibitem{jin2022gnnlens}
Z.~Jin, Y.~Wang, Q.~Wang, Y.~Ming, T.~Ma, and H.~Qu.
\newblock {GNNLens}: A visual analytics approach for prediction error diagnosis of graph neural networks.
\newblock {\em IEEE Transactions on Visualization and Computer Graphics}, 29(6):3024--3038, 2022. \href{https://doi.org/10.1109/TVCG.2022.3148107}
{doi: {{%
10\hspace{.1pt}\discretionary{.}{%
}{.}\hspace{.4pt}1109\discretionary{/}{%
}{/}TVCG\hspace{.1pt}\discretionary{.}{%
}{.}\hspace{.4pt}2022\hspace{.1pt}\discretionary{.}{%
}{.}\hspace{.4pt}3148107}}}


\bibitem{langgraph}
{LangChain}.
\newblock {LangGraph. Building language agents as graphs.}

\bibitem{langgraphstudio2024}
LangChain.
\newblock {LangGraph Studio: The first agent IDE}.
\newblock \url{https://blog.langchain.dev/langgraph-studio-the-first-agent-ide/}.

\bibitem{langgraph2024multiagent}
LangChain.
\newblock {Multi-agent Systems}.
\newblock \url{https://langchain-ai.github.io/langgraph/concepts/multi_agent/}.

\bibitem{langgraph2024workflows}
LangChain.
\newblock {Multi-agent Workflows}.
\newblock \url{https://langchain-ai.github.io/langgraph/tutorials/workflows/}.

\bibitem{li2023camel}
G.~Li, H.~A. Al~Kader~Hammoud, H.~Itani, D.~Khizbullin, and B.~Ghanem.
\newblock {CAMEL: Communicative Agents for "Mind" Exploration of Large Language Model Society}.
\newblock In {\em Proceedings of the 37th International Conference on Neural Information Processing Systems}, NIPS '23,  article no. 2264,  18 pages. Curran Associates Inc., Red Hook, 2023. \href{https://doi.org/10.5555/3666122.3668386}
{doi: {{%
10\hspace{.1pt}\discretionary{.}{%
}{.}\hspace{.4pt}5555\discretionary{/}{%
}{/}3666122\hspace{.1pt}\discretionary{.}{%
}{.}\hspace{.4pt}3668386}}}


\bibitem{liang2024debatedivergent}
T.~Liang, Z.~He, W.~Jiao, X.~Wang, Y.~Wang, R.~Wang, Y.~Yang, S.~Shi, and Z.~Tu.
\newblock {Encouraging Divergent Thinking in Large Language Models through Multi-Agent Debate}.
\newblock In Y.~Al-Onaizan, M.~Bansal, and Y.-N. Chen, eds., {\em Proceedings of the 2024 Conference on Empirical Methods in Natural Language Processing}, pp. 17889--17904. Association for Computational Linguistics, Miami, Florida, USA, 2024. \href{https://doi.org/10.18653/v1/2024.emnlp-main.992}
{doi: {{%
10\hspace{.1pt}\discretionary{.}{%
}{.}\hspace{.4pt}18653\discretionary{/}{%
}{/}v1\discretionary{/}{%
}{/}2024\hspace{.1pt}\discretionary{.}{%
}{.}\hspace{.4pt}emnlp\discretionary{%
}{-}{-}main\hspace{.1pt}\discretionary{.}{%
}{.}\hspace{.4pt}992}}}


\bibitem{liu2016towards}
M.~Liu, J.~Shi, Z.~Li, C.~Li, J.~Zhu, and S.~Liu.
\newblock Towards better analysis of deep convolutional neural networks.
\newblock {\em IEEE Transactions on Visualization and Computer Graphics}, 23(1):91--100, 2016. \href{https://doi.org/10.1109/TVCG.2016.2598831}
{doi: {{%
10\hspace{.1pt}\discretionary{.}{%
}{.}\hspace{.4pt}1109\discretionary{/}{%
}{/}TVCG\hspace{.1pt}\discretionary{.}{%
}{.}\hspace{.4pt}2016\hspace{.1pt}\discretionary{.}{%
}{.}\hspace{.4pt}2598831}}}


\bibitem{lu2024gnn}
Y.~Lu, C.~Chen, Y.~Chen, K.~Huang, M.~Zitnik, and Q.~Wang.
\newblock Gnn 101: Visual learning of graph neural networks in your web browser.
\newblock {\em arXiv preprint arXiv:2411.17849}, 2024. \href{https://doi.org/10.48550/arXiv.2411.17849}
{doi: {{%
10\hspace{.1pt}\discretionary{.}{%
}{.}\hspace{.4pt}48550\discretionary{/}{%
}{/}arXiv\hspace{.1pt}\discretionary{.}{%
}{.}\hspace{.4pt}2411\hspace{.1pt}\discretionary{.}{%
}{.}\hspace{.4pt}17849}}}


\bibitem{microsoft2024conversation}
{Microsoft AutoGen}.
\newblock {Conversation Patterns}.
\newblock \url{https://microsoft.github.io/autogen/0.2/docs/tutorial/conversation-patterns/}.

\bibitem{microsoft2024patterns}
{Microsoft AutoGen}.
\newblock {Multi-Agent Design Patteamrns}.
\newblock \url{https://microsoft.github.io/autogen/stable/user-guide/core-user-guide/design-patterns/intro.html}.

\bibitem{munzner2009nested}
T.~Munzner.
\newblock A nested model for visualization design and validation.
\newblock {\em IEEE Transactions on Visualization and Computer Graphics}, 15(6):921--928, 2009.

\bibitem{nakano2022webgptbrowser}
R.~Nakano, J.~Hilton, S.~Balaji, J.~Wu, L.~Ouyang, C.~Kim, C.~Hesse, S.~Jain, V.~Kosaraju, W.~Saunders, X.~Jiang, K.~Cobbe, T.~Eloundou, G.~Krueger, K.~Button, M.~Knight, B.~Chess, and J.~Schulman.
\newblock {WebGPT: Browser-assisted question-answering with human feedback}, 2022. \href{https://doi.org/10.48550/arXiv.2112.09332}
{doi: {{%
10\hspace{.1pt}\discretionary{.}{%
}{.}\hspace{.4pt}48550\discretionary{/}{%
}{/}arXiv\hspace{.1pt}\discretionary{.}{%
}{.}\hspace{.4pt}2112\hspace{.1pt}\discretionary{.}{%
}{.}\hspace{.4pt}09332}}}


\bibitem{niu2025flowmodular}
B.~Niu, Y.~Song, K.~Lian, Y.~Shen, Y.~Yao, K.~Zhang, and T.~Liu.
\newblock {Flow: Modularized Agentic Workflow Automation}, 2025. \href{https://doi.org/10.48550/arXiv.2501.07834}
{doi: {{%
10\hspace{.1pt}\discretionary{.}{%
}{.}\hspace{.4pt}48550\discretionary{/}{%
}{/}arXiv\hspace{.1pt}\discretionary{.}{%
}{.}\hspace{.4pt}2501\hspace{.1pt}\discretionary{.}{%
}{.}\hspace{.4pt}07834}}}


\bibitem{openai2024gpt4ocard}
{OpenAI}.
\newblock {GPT-4o System Card}, 2024. \href{https://doi.org/10.48550/arXiv.2410.21276}
{doi: {{%
10\hspace{.1pt}\discretionary{.}{%
}{.}\hspace{.4pt}48550\discretionary{/}{%
}{/}arXiv\hspace{.1pt}\discretionary{.}{%
}{.}\hspace{.4pt}2410\hspace{.1pt}\discretionary{.}{%
}{.}\hspace{.4pt}21276}}}


\bibitem{park2023generative}
J.~S. Park, J.~O'Brien, C.~J. Cai, M.~R. Morris, P.~Liang, and M.~S. Bernstein.
\newblock {Generative Agents: Interactive Simulacra of Human Behavior}.
\newblock In {\em Proceedings of the 36th Annual ACM Symposium on User Interface Software and Technology}, UIST '23,  article no. 2,  22 pages. ACM, New York, 2023. \href{https://doi.org/10.1145/3586183.3606763}
{doi: {{%
10\hspace{.1pt}\discretionary{.}{%
}{.}\hspace{.4pt}1145\discretionary{/}{%
}{/}3586183\hspace{.1pt}\discretionary{.}{%
}{.}\hspace{.4pt}3606763}}}


\bibitem{penades2025ai}
J.~R. Penadés, J.~Gottweis, L.~He, J.~B. Patkowski, A.~Shurick, W.-H. Weng, T.~Tu, A.~Palepu, A.~Myaskovsky, A.~Pawlosky, N.~Vivek, K.~Alan, and T.~R.~D. Costa.
\newblock {AI mirrors experimental science to uncover a novel mechanism of gene transfer crucial to bacterial evolution}.
\newblock {\em bioRxiv}, pp. 2025--02, 2025. \href{https://doi.org/10.1101/2025.02.19.639094}
{doi: {{%
10\hspace{.1pt}\discretionary{.}{%
}{.}\hspace{.4pt}1101\discretionary{/}{%
}{/}2025\hspace{.1pt}\discretionary{.}{%
}{.}\hspace{.4pt}02\hspace{.1pt}\discretionary{.}{%
}{.}\hspace{.4pt}19\hspace{.1pt}\discretionary{.}{%
}{.}\hspace{.4pt}639094}}}


\bibitem{prasad2023adapt}
A.~Prasad, A.~Koller, M.~Hartmann, P.~Clark, A.~Sabharwal, M.~Bansal, and T.~Khot.
\newblock {ADaPT: As-Needed Decomposition and Planning with Language Models}.
\newblock In K.~Duh, H.~Gomez, and S.~Bethard, eds., {\em Findings of the Association for Computational Linguistics: NAACL 2024}, pp. 4226--4252. Association for Computational Linguistics, Mexico City, 2024. \href{https://doi.org/10.18653/v1/2024.findings-naacl.264}
{doi: {{%
10\hspace{.1pt}\discretionary{.}{%
}{.}\hspace{.4pt}18653\discretionary{/}{%
}{/}v1\discretionary{/}{%
}{/}2024\hspace{.1pt}\discretionary{.}{%
}{.}\hspace{.4pt}findings\discretionary{%
}{-}{-}naacl\hspace{.1pt}\discretionary{.}{%
}{.}\hspace{.4pt}264}}}


\bibitem{qian2024chatdev}
C.~Qian, W.~Liu, H.~Liu, N.~Chen, Y.~Dang, J.~Li, C.~Yang, W.~Chen, Y.~Su, X.~Cong, J.~Xu, D.~Li, Z.~Liu, and M.~Sun.
\newblock {ChatDev: Communicative Agents for Software Development}.
\newblock In L.-W. Ku, A.~Martins, and V.~Srikumar, eds., {\em Proceedings of the 62nd Annual Meeting of the Association for Computational Linguistics (Volume 1: Long Papers)}, pp. 15174--15186. Association for Computational Linguistics, Bangkok, Thailand, 2024. \href{https://doi.org/10.18653/v1/2024.acl-long.810}
{doi: {{%
10\hspace{.1pt}\discretionary{.}{%
}{.}\hspace{.4pt}18653\discretionary{/}{%
}{/}v1\discretionary{/}{%
}{/}2024\hspace{.1pt}\discretionary{.}{%
}{.}\hspace{.4pt}acl\discretionary{%
}{-}{-}long\hspace{.1pt}\discretionary{.}{%
}{.}\hspace{.4pt}810}}}


\bibitem{shinn2023reflexion}
N.~Shinn, F.~Cassano, E.~Berman, A.~Gopinath, K.~Narasimhan, and S.~Yao.
\newblock {Reflexion: Language Agents with Verbal Reinforcement Learning}, 2023. \href{https://doi.org/10.48550/arXiv.2303.11366}
{doi: {{%
10\hspace{.1pt}\discretionary{.}{%
}{.}\hspace{.4pt}48550\discretionary{/}{%
}{/}arXiv\hspace{.1pt}\discretionary{.}{%
}{.}\hspace{.4pt}2303\hspace{.1pt}\discretionary{.}{%
}{.}\hspace{.4pt}11366}}}


\bibitem{suh2024luminate}
S.~Suh, M.~Chen, B.~Min, T.~J.-J. Li, and H.~Xia.
\newblock {Luminate: Structured Generation and Exploration of Design Space with Large Language Models for Human-AI Co-Creation}.
\newblock In {\em Proceedings of the 2024 CHI Conference on Human Factors in Computing Systems}, CHI '24,  article no. 644,  26 pages. ACM, New York, 2024. \href{https://doi.org/10.1145/3613904.3642400}
{doi: {{%
10\hspace{.1pt}\discretionary{.}{%
}{.}\hspace{.4pt}1145\discretionary{/}{%
}{/}3613904\hspace{.1pt}\discretionary{.}{%
}{.}\hspace{.4pt}3642400}}}


\bibitem{suh2023sensecape}
S.~Suh, B.~Min, S.~Palani, and H.~Xia.
\newblock {Sensecape: Enabling Multilevel Exploration and Sensemaking with Large Language Models}.
\newblock In {\em Proceedings of the 36th Annual ACM Symposium on User Interface Software and Technology}, UIST '23,  article no. 1,  18 pages. ACM, New York, 2023. \href{https://doi.org/10.1145/3586183.3606756}
{doi: {{%
10\hspace{.1pt}\discretionary{.}{%
}{.}\hspace{.4pt}1145\discretionary{/}{%
}{/}3586183\hspace{.1pt}\discretionary{.}{%
}{.}\hspace{.4pt}3606756}}}


\bibitem{vegaLiteMovies}
{Vega-Lite}.
\newblock Movies dataset.
\newblock \url{https://vega.github.io/vega-lite/data/movies.json}, 2017.

\bibitem{AG2_2024}
C.~Wang, Q.~Wu, and the AG2~Community.
\newblock {AG2: Open-Source AgentOS for AI Agents}.
\newblock \url{https://docs.ag2.ai/}, 2024.
\newblock Available at https://docs.ag2.ai/.

\bibitem{wang2024surveyagents}
L.~Wang, C.~Ma, X.~Feng, Z.~Zhang, H.~Yang, J.~Zhang, Z.~Chen, J.~Tang, X.~Chen, Y.~Lin, W.~X. Zhao, Z.~Wei, and J.~Wen.
\newblock {A survey on large language model based autonomous agents}.
\newblock {\em Frontiers of Computer Science}, 18(6), 2024. \href{https://doi.org/10.1007/s11704-024-40231-1}
{doi: {{%
10\hspace{.1pt}\discretionary{.}{%
}{.}\hspace{.4pt}1007\discretionary{/}{%
}{/}s11704\discretionary{%
}{-}{-}024\discretionary{%
}{-}{-}40231\discretionary{%
}{-}{-}1}}}


\bibitem{wang2022extending}
Q.~Wang, K.~Huang, P.~Chandak, M.~Zitnik, and N.~Gehlenborg.
\newblock Extending the nested model for user-centric xai: A design study on gnn-based drug repurposing.
\newblock {\em IEEE Transactions on Visualization and Computer Graphics}, 29(1):1266--1276, 2022. \href{https://doi.org/10.1109/TVCG.2022.3209435}
{doi: {{%
10\hspace{.1pt}\discretionary{.}{%
}{.}\hspace{.4pt}1109\discretionary{/}{%
}{/}TVCG\hspace{.1pt}\discretionary{.}{%
}{.}\hspace{.4pt}2022\hspace{.1pt}\discretionary{.}{%
}{.}\hspace{.4pt}3209435}}}


\bibitem{wang2019atmseer}
Q.~Wang, Y.~Ming, Z.~Jin, Q.~Shen, D.~Liu, M.~J. Smith, K.~Veeramachaneni, and H.~Qu.
\newblock {ATMSeer: Increasing Transparency and Controllability in Automated Machine Learning}.
\newblock In {\em Proceedings of the 2019 CHI Conference on Human Factors in Computing Systems}, CHI '19,  12 pages, p. 1–12. ACM, New York, 2019. \href{https://doi.org/10.1145/3290605.3300911}
{doi: {{%
10\hspace{.1pt}\discretionary{.}{%
}{.}\hspace{.4pt}1145\discretionary{/}{%
}{/}3290605\hspace{.1pt}\discretionary{.}{%
}{.}\hspace{.4pt}3300911}}}


\bibitem{wang2019visual}
Q.~Wang, J.~Yuan, S.~Chen, H.~Su, H.~Qu, and S.~Liu.
\newblock {Visual Genealogy of Deep Neural Networks}.
\newblock {\em IEEE Transactions on Visualization and Computer Graphics}, 26(11):3340--3352, 2020. \href{https://doi.org/10.1109/TVCG.2019.2921323}
{doi: {{%
10\hspace{.1pt}\discretionary{.}{%
}{.}\hspace{.4pt}1109\discretionary{/}{%
}{/}TVCG\hspace{.1pt}\discretionary{.}{%
}{.}\hspace{.4pt}2019\hspace{.1pt}\discretionary{.}{%
}{.}\hspace{.4pt}2921323}}}


\bibitem{wang2024openworld}
Z.~Wang, S.~Cai, G.~Chen, A.~Liu, X.~Ma, Y.~Liang, and T.~CraftJarvis.
\newblock {Describe, Explain, Plan and Select: Interactive Planning with Large Language Models Enables Open-World Multi-Task Agents}.
\newblock In {\em Proceedings of the 37th International Conference on Neural Information Processing Systems}, NIPS '23,  article no. 1480,  37 pages. Curran Associates Inc., Red Hook, 2023.

\bibitem{wu2023autogen}
Q.~Wu, G.~Bansal, J.~Zhang, Y.~Wu, B.~Li, E.~Zhu, L.~Jiang, X.~Zhang, S.~Zhang, J.~Liu, et~al.
\newblock {AutoGen: Enabling next-gen llm applications via multi-agent conversation}.
\newblock {\em arXiv preprint arXiv:2308.08155}, 2023. \href{https://doi.org/10.48550/arXiv.2308.08155}
{doi: {{%
10\hspace{.1pt}\discretionary{.}{%
}{.}\hspace{.4pt}48550\discretionary{/}{%
}{/}arXiv\hspace{.1pt}\discretionary{.}{%
}{.}\hspace{.4pt}2308\hspace{.1pt}\discretionary{.}{%
}{.}\hspace{.4pt}08155}}}


\bibitem{wu2022prompt}
T.~Wu, E.~Jiang, A.~Donsbach, J.~Gray, A.~Molina, M.~Terry, and C.~J. Cai.
\newblock {PromptChainer: Chaining Large Language Model Prompts through Visual Programming}.
\newblock In {\em Extended Abstracts of the 2022 CHI Conference on Human Factors in Computing Systems}, CHI EA '22,  article no. 359,  10 pages. ACM, New York, 2022. \href{https://doi.org/10.1145/3491101.3519729}
{doi: {{%
10\hspace{.1pt}\discretionary{.}{%
}{.}\hspace{.4pt}1145\discretionary{/}{%
}{/}3491101\hspace{.1pt}\discretionary{.}{%
}{.}\hspace{.4pt}3519729}}}


\bibitem{wu2022chains}
T.~Wu, M.~Terry, and C.~J. Cai.
\newblock {AI Chains: Transparent and Controllable Human-AI Interaction by Chaining Large Language Model Prompts}.
\newblock In {\em Proceedings of the 2022 CHI Conference on Human Factors in Computing Systems}, CHI '22,  article no. 385,  22 pages. ACM, New York, 2022. \href{https://doi.org/10.1145/3491102.3517582}
{doi: {{%
10\hspace{.1pt}\discretionary{.}{%
}{.}\hspace{.4pt}1145\discretionary{/}{%
}{/}3491102\hspace{.1pt}\discretionary{.}{%
}{.}\hspace{.4pt}3517582}}}


\bibitem{wu2023crowdsoucingllms}
T.~Wu, H.~Zhu, M.~Albayrak, A.~Axon, A.~Bertsch, W.~Deng, Z.~Ding, B.~Guo, S.~Gururaja, T.-S. Kuo, J.~T. Liang, R.~Liu, I.~Mandal, J.~Milbauer, X.~Ni, N.~Padmanabhan, S.~Ramkumar, A.~Sudjianto, J.~Taylor, Y.-J. Tseng, P.~Vaidos, Z.~Wu, W.~Wu, and C.~Yang.
\newblock {LLMs as Workers in Human-Computational Algorithms? Replicating Crowdsourcing Pipelines with LLMs}.
\newblock In {\em Proceedings of the Extended Abstracts of the CHI Conference on Human Factors in Computing Systems}, CHI EA '25,  article no. 684,  10 pages. ACM, New York, 2025. \href{https://doi.org/10.1145/3706599.3706690}
{doi: {{%
10\hspace{.1pt}\discretionary{.}{%
}{.}\hspace{.4pt}1145\discretionary{/}{%
}{/}3706599\hspace{.1pt}\discretionary{.}{%
}{.}\hspace{.4pt}3706690}}}


\bibitem{yuan2022visual}
J.~Yuan, M.~Liu, F.~Tian, and S.~Liu.
\newblock {Visual Analysis of Neural Architecture Spaces for Summarizing Design Principles}.
\newblock {\em IEEE Transactions on Visualization and Computer Graphics}, 29(1):288--298, 2023. \href{https://doi.org/10.1109/TVCG.2022.3209404}
{doi: {{%
10\hspace{.1pt}\discretionary{.}{%
}{.}\hspace{.4pt}1109\discretionary{/}{%
}{/}TVCG\hspace{.1pt}\discretionary{.}{%
}{.}\hspace{.4pt}2022\hspace{.1pt}\discretionary{.}{%
}{.}\hspace{.4pt}3209404}}}


\bibitem{liu2024cataloque}
L.~Yue, L.~Sin~Kit, L.~Qinghua, Z.~Liming, Z.~Dehai, X.~Xiwei, H.~Stefan, and W.~Jon.
\newblock {Agent Design Pattern Catalogue: A Collection of Architectural Patterns for Foundation Model based Agents}.
\newblock {\em arXiv preprint arXiv: 2405.10467}, 2024. \href{https://doi.org/10.48550/arXiv.2405.10467}
{doi: {{%
10\hspace{.1pt}\discretionary{.}{%
}{.}\hspace{.4pt}48550\discretionary{/}{%
}{/}arXiv\hspace{.1pt}\discretionary{.}{%
}{.}\hspace{.4pt}2405\hspace{.1pt}\discretionary{.}{%
}{.}\hspace{.4pt}10467}}}


\bibitem{zhang2025chainbuddy}
J.~Zhang and I.~Arawjo.
\newblock {ChainBuddy: An AI-assisted Agent System for Helping Users Set up LLM Pipelines}.
\newblock In {\em Adjunct Proceedings of the 37th Annual ACM Symposium on User Interface Software and Technology}, UIST Adjunct '24,  article no. 48,  3 pages. ACM, New York, 2024. \href{https://doi.org/10.1145/3672539.3686763}
{doi: {{%
10\hspace{.1pt}\discretionary{.}{%
}{.}\hspace{.4pt}1145\discretionary{/}{%
}{/}3672539\hspace{.1pt}\discretionary{.}{%
}{.}\hspace{.4pt}3686763}}}


\bibitem{zhang2024aflow}
J.~Zhang, J.~Xiang, Z.~Yu, F.~Teng, X.-H. Chen, J.~Chen, M.~Zhuge, X.~Cheng, S.~Hong, J.~Wang, B.~Zheng, B.~Liu, Y.~Luo, and C.~Wu.
\newblock {AFlow: Automating Agentic Workflow Generation}.
\newblock In {\em The Thirteenth International Conference on Learning Representations}, 2025. \href{https://doi.org/10.48550/arXiv.2410.10762}
{doi: {{%
10\hspace{.1pt}\discretionary{.}{%
}{.}\hspace{.4pt}48550\discretionary{/}{%
}{/}arXiv\hspace{.1pt}\discretionary{.}{%
}{.}\hspace{.4pt}2410\hspace{.1pt}\discretionary{.}{%
}{.}\hspace{.4pt}10762}}}


\bibitem{zhang2024optimize}
S.~Zhang, J.~Zhang, J.~Liu, L.~Song, C.~Wang, R.~Krishna, and Q.~Wu.
\newblock {Offline Training of Language Model Agents with Functions as Learnable Weights}.
\newblock In {\em Proceedings of the 41st International Conference on Machine Learning}, ICML'24,  article no. 2496,  21 pages. JMLR.org, 2024.

\bibitem{zheng2023judging}
L.~Zheng, W.-L. Chiang, Y.~Sheng, S.~Zhuang, Z.~Wu, Y.~Zhuang, Z.~Lin, Z.~Li, D.~Li, E.~Xing, et~al.
\newblock Judging {LLM-as-a-Judge} with {MT-Bench} and chatbot arena.
\newblock {\em Advances in Neural Information Processing Systems}, 36:46595--46623, 2023.

\bibitem{zhou2025multi}
H.~Zhou, X.~Wan, R.~Sun, H.~Palangi, S.~Iqbal, I.~Vuli{\'c}, A.~Korhonen, and S.~{\"O}. Ar{\i}k.
\newblock {Multi-Agent Design: Optimizing Agents with Better Prompts and Topologies}.
\newblock {\em arXiv preprint arXiv:2502.02533}, 2025. \href{https://doi.org/10.48550/arXiv.2502.02533}
{doi: {{%
10\hspace{.1pt}\discretionary{.}{%
}{.}\hspace{.4pt}48550\discretionary{/}{%
}{/}arXiv\hspace{.1pt}\discretionary{.}{%
}{.}\hspace{.4pt}2502\hspace{.1pt}\discretionary{.}{%
}{.}\hspace{.4pt}02533}}}


\bibitem{zhuge2024gptswarm}
M.~Zhuge, W.~Wang, L.~Kirsch, F.~Faccio, D.~Khizbullin, and J.~Schmidhuber.
\newblock {GPTSwarm: Language Agents as Optimizable Graphs}.
\newblock In R.~Salakhutdinov, Z.~Kolter, K.~Heller, A.~Weller, N.~Oliver, J.~Scarlett, and F.~Berkenkamp, eds., {\em Proceedings of the 41st International Conference on Machine Learning}, vol. 235 of {\em Proceedings of Machine Learning Research}, pp. 62743--62767. PMLR, 2024. \href{https://doi.org/10.48550/arXiv.2402.16823}
{doi: {{%
10\hspace{.1pt}\discretionary{.}{%
}{.}\hspace{.4pt}48550\discretionary{/}{%
}{/}arXiv\hspace{.1pt}\discretionary{.}{%
}{.}\hspace{.4pt}2402\hspace{.1pt}\discretionary{.}{%
}{.}\hspace{.4pt}16823}}}


\end{thebibliography}








\end{document}